\documentclass[aps,prd,preprint,a4paper,showpacs,nofootinbib,superscriptaddress]{revtex4-2}
\usepackage{bm}
\usepackage{indentfirst}
\usepackage{amsmath}
\usepackage{graphicx}

\usepackage{amssymb}
\usepackage{subfigure}
\usepackage{amssymb}
\usepackage{hyperref}
\usepackage{epstopdf}
\usepackage[section]{placeins}

\usepackage[utf8]{inputenc}
\hypersetup{
    colorlinks=true,
    linkcolor=red,
    citecolor=blue,
}
\usepackage{color}
\usepackage[T1]{fontenc}
\usepackage{txfonts}
\begin{document}

\title{Detecting the secondary spin with extreme mass ratio inspirals in Scalar-Tensor theory
%Detecting secondary spin around Kerr black hole with scalar field through Extreme Mass Ratio Inspirals
}

\author{Hong Guo}
\email{gravhguo@gmail.com}
\affiliation{\textit{Center for Gravitation and Cosmology, College of Physical
Science and Technology, Yangzhou University, Yangzhou 225009, China}}
%\affiliation{\textit{School of Physics and Astronomy, Shanghai Jiao Tong
%University, Shanghai 200240, China}}

\author{Chao Zhang}
\email{zhangchao666@sjtu.edu.cn}
\affiliation{\textit{Shanghai Frontier Research Center for Gravitational Wave Detection, Shanghai Jiao Tong University, Shanghai 200240, China}}

\author{Yunqi Liu}
\email{yunqiliu@yzu.edu.cn (corresponding author)}
\affiliation{\textit{Center for Gravitation and Cosmology, College of Physical
Science and Technology, Yangzhou University, Yangzhou 225009, China}}
\affiliation{\textit{School of Aeronautics and Astronautics, Shanghai Jiao Tong
University, Shanghai 200240, China}}

\author{Rui-Hong Yue}
\email{rhyue@yzu.edu.cn}
\affiliation{\textit{Center for Gravitation and Cosmology, College of Physical Science and Technology, Yangzhou University, Yangzhou 225009, China}}

\author{Yungui Gong}
\email{yggong@hust.edu.cn}
\affiliation{School of Physics, Huazhong University of Science and Technology, Wuhan, Hubei 430074, China}

\author{Bin Wang}
\email{wang_b@sjtu.edu.cn}
\affiliation{\textit{Center for Gravitation and Cosmology, College of Physical Science and Technology, Yangzhou University, Yangzhou 225009, China}}
\affiliation{\textit{School of Aeronautics and Astronautics, Shanghai Jiao Tong University, Shanghai 200240, China}}

\begin{abstract}
In this paper, we investigate the detectability of secondary spin in the extreme mass ratio inspirals(EMRI) system within a modified gravity model coupled with a scalar field. 
The central black hole, which reduces to a Kerr one, is circularly spiralled by a scalar-charged spinning secondary body on the equatorial plane.  
The analysis reveals that the presence of the scalar field amplifies the secondary spin effect, allowing for a lower limit of the detectability and an improved resolution of the secondary spin when the scalar charge is sufficiently large.
Our findings suggest that the secondary spin detection is more feasible when the primary mass is not large, and TianQin is the optimal choice for detection.
%Furthermore, our study provides a modified gravity EMRI model that can be utilized to examine binary properties in these theories, which is an important direction for future research given the potential of alternative gravity theories for studying the universe. 
\end{abstract}

\maketitle

\newpage

%%%%%%%%%%%%%%%%%%%%%%%%%%%%%%%%%%%%%%%%%%%%%%%%%%%%%%%%%%%%%%%%%%%%%%%%%%%%
\section{Introduction}\label{sec=intro}

The extreme-gravity regions of the universe, where black holes (BHs) and compact objects (COs) reside, are a treasure trove for testing the theory of gravity and exploring the secrets of spacetime.
The detection of gravitational waves (GWs)  \cite{LIGOScientific:2016vlm,LIGOScientific:2018mvr,LIGOScientific:2019fpa} has opened new channels to probe these highly-dynamical, strong-curvature regions by observing binaries of BHs and COs \cite{Yagi:2016jml,Berti:2018cxi,Barack:2018yly}.
The study of GW astrophysics inspires us with great confidence in fundamental theories and new physics \cite{Belenchia:2021rfb,Vagie:2021bup}.
One of the important frequency windows for GW detections is the extreme mass ratio inspiral(EMRI) \cite{Berry:2019wgg}.
During the inspirals of a stellar mass body (the secondary) orbiting a central supermassive black hole (the primary), the EMRI system radiates tens to hundreds of thousands of GW cycles.
The accumulated signal provides an effective tool to probe the near-horizon environment of black holes \cite{Yunes:2011ws,Kocsis:2011dr,Derdzinski:2020wlw, Hannuksela:2019vip}, source localization \cite{Eracleous:2019bal}, massive BH spectrum and corresponding evolution \cite{Babak:2017tow,saglia:2016sinfoni,Gair:2010yu}.
Space-based GW detectors that are currently under development, such as the Laser Interferometer Space Antenna (LISA) \cite{LISA:2017pwj,Danzmann:1997hm}, DECIGO \cite{Kawamura:2006up}, Taiji \cite{Hu:2017mde,Gong:2021gvw} and TianQin \cite{TianQin:2015yph,Gong:2021gvw}, will target EMRIs as their primary detection sources.
%These detectors will be crucial for advancing GW physics and related astronomy \cite{Carson:2019yxq,Bailes:2021tot,Foucart:2022iwu} in the coming decades.

%The primary supermassive BHs in EMRIs are mostly located at the center of galactic nuclei \cite{Han:2020dql,Yang:2022vml} which are characterized by the extremely complex matter environment including dark matter \cite{Hannuksela:2019vip}, Accretion Disk \cite{Yunes:2011ws,Kocsis:2011dr} and so on \cite{Cardoso:2021wlq}.
%The nearby spacetime of the primary BH is affected by any stress-energy tensor, which modifies the inspiral of the secondary, and leaves imprints in the EMRI waveform. 
%Recently, reference \cite{Cardoso:2022whc} presented an EMRI model of a BH within a halo of matter using a generic fully-relativistic formalism and analyzed the detection of galactic properties.
%The effects of accretion-disk effects \cite{Speri:2022upm} and dark matter halo \cite{Destounis:2022obl,Dai:2023cft,Figueiredo:2023gas} on EMRI systems have also been studied. 

In general relativity (GR), binary systems emit GWs with tensor polarizations, and the lowest radiative multipole moment is the quadrupole moment.
However, in alternative theories of gravity, additional emission channels may exist.
For example, in Brans-Dicke theory and some scalar-tensor theories, the additional
scalar field activates the dipole gravitational radiation reaction \cite{Alsing:2011er,PhysRevD.56.785,PhysRevD.66.024040,Kuntz_2020}.
Even if these additional modes physically exist, the milli-Hz low frequency band GWs have significantly lower strength compared to the tensorial polarizations\cite{dePaula:2004bc,Hou:2017bqj}, making them more difficult to observe by the planned space-borne missions.
The characteristics of the long-time duration of EMRIs make it a competitive way to detect these additional modes.
Typically, the entire process of the inspiral of the secondary in EMRIs lasts for tens to hundreds of years.
As a result, although the instantaneous strength of the additional radiation is less significant when compared with the tensor modes, the accumulated signal could possibly deviate from that determined by GR, with the dynamics of the EMRI systems modified by the presence of the additional radiation.
In Ref.\cite{Maselli:2020zgv}, the authors showed that for specific classes of scalar-tensor alternative gravity theories, EMRI could be considered as a test particle with a scalar charge inspiraling onto the central supermassive BH.
%\red{This scalar charge, proportional to the nonminimal coupling of the theory between the metric field and the scalar field, is not related to any conservative quantity}.
They demonstrated that the corresponding dephasing caused by the scalar radiation should be detectable by LISA. 
This is an impressive model that offers twofold benefits.
On the one hand, one allows for the study of extra scalar radiation and how it differs from EMRI waveforms in GR, providing a way to test gravity theories in the strong field regime.
Ref.\cite{Maselli:2021men} studied LISA's ability to detect the model-independent scalar charge and proposed a GW template for detecting new fundamental fields in our universe.
The research has been extended to Kerr spacetime \cite{Guo:2022euk}, eccentric equatorial orbits \cite{Barsanti:2022ana,Zhang:2022rfr}, and the massive scalar field case \cite{Barsanti:2022vvl}.
Additionally, in Ref.\cite{Zhang:2022hbt}, the extra radiation of the electromagnetic field was considered for the first time, and the detectability of the electromagnetic charge by EMRI GW signals from LISA was analyzed. Several recent studies have focused on this topic \cite{Liang:2022gdk,Zhang:2023vok}.
On the other hand, it puts forward an EMRI model which allows for investigation into the properties of binaries in these modified gravity.
One particular property of interest is the effect of secondary spin during inspirals, which has been extensively studied in EMRIs \cite{Huerta:2011kt,Chicone:2005jj,Singh:2008qr}.

Many astrophysically relevant BHs or COs have nonzero angular momentum\cite{Burko:2003rv,Hartl:2002ig}, to obtain high accuracy theoretical waveform it is reasonable to take into account the spin of the secondary in EMRIs \cite{Huerta:2011kt,Akcay:2019bvk,Witzany:2019dii}. 
Furthermore, precise detection of the secondary's spin can help us study the properties of the secondary objects, which is an initial step in building a spectrum of stellar-mass to intermediate-black-hole-mass compact objects \cite{Babak:2017tow}. 
However, most studies on the effect of the secondary spin in EMRIs have been limited to GR (see recent works \cite{Piovano:2020ooe,Piovano:2020zin,Rahman:2021eay,Skoupy:2022adh,Drummond:2022xej}), it would be valuable to extend this research to modified gravity.
Firstly, the spin-curvature interaction will deviate the secondary from geodesic motion in GR. Then, the secondary spin contributions may also arise from the additional radiation of the GW signal in the modified gravitational EMRI model. With these considerations, the extra scalar energy fluxes potentially enhance the secondary spin effect in GWs, resulting in the improvement of secondary spin detection by space-based GW detectors.
Despite this potential, none of the previous work related to the model in Ref.\cite{Maselli:2020zgv} has computed the spin-correction of the secondary to the GW phase. It is interesting to focus our attention on this topic.

In the present paper, our objective is to extend the model proposed in Ref.\cite{Maselli:2020zgv} to study the detection of  spin-corrections of the secondary.
The spin-curvature coupling, described by the Mathisson-Papapetrou-Dixon (MPD) equations, is the leading order effect of the finite size of a rapidly rotating compact astrophysical object moving in a curved background. 
It is a next-to-leading order effect in the phase of GWs emitted by EMRIs, and is expected to be comparable to the effect induced by the additional scalar radiation.
Therefore, as one aims to detect the scalar charge by studying the additional scalar radiation, it is reasonable to wish for secondary spin detection with the additional scalar radiation.
In this EMRI model, we calculate the related GW fluxes and GW phases and compare the detectability of LISA, Taiji and TianQin using the parameter estimation approaches.
The results demonstrate that the presence of the scalar field amplifies the secondary spin effect, allowing for the detection of a lower limit value of secondary spin, and an improved resolution of secondary spin detection when the scalar charge is sufficiently large.

%Before delving into the main text, it is necessary to introduce the data processing approach employed in this paper.
%Different from the original approach considering the one-year evolution prior to the secondary inspiraling into the innermost stable circular orbit (ISCO), we prefer to start the one-year evolution from the same initial location and ensure that the secondary is terminated outside the ISCO but as close as possible to the ISCO taking into account all the considered parameters.
%This modified approach has been widely used in the study of secondary spin by EMRIs \cite{Piovano:2020ooe,Piovano:2020zin,Rahman:2021eay,Skoupy:2022adh,Drummond:2022xej}.
%While the modified method adds an extra constraint on the initial position of the secondary, it considerably improves the resolution and accuracy of dephasing.
 
The paper is organized as follows: Sec.\ref{sec=model} constructs the EMRI system and introduces the orbit motion of the spinning test particle in the Kerr spacetime. 
Sec.\ref{sec=perturb} gives the tensor perturbation and scalar perturbation respectively and the energy fluxes are obtained by the solution of the perturbation equations.
In Sec.\ref{sec=result} we discuss the orbital evolution, total energy fluxes, dephasing and related faithfulness of the GW signals.
Finally, we summarize our results and present concluding remarks in Sec.\ref{sec=conclusion}. Throughout the paper, we use geometric units as $c=G=1.$

%%%%%%%%%%%%%%%%%%%%%%%%%%%%%%%%%%%%%%%%%%%%%%%%%%%%%%%%%%%%%%%%%%%%%%%%%%%%
\section{Model construction}\label{sec=model}

In this section, we examine the case of a spinning secondary spiraling into a central supermassive Kerr BH with the inclusion of a scalar field coupling to higher-order curvature instants during the quasi-circular orbital evolution in the equatorial plane. 
We will begin by outlining the model framework and then provide a brief overview of the orbital motion of the spinning particle.

%%%%%%%%%%%%%%%%%%%%%%%%%%%%%%%%%%%%%%%%%%%%%%%%%%%%%%%%%%%%%%%%%%%%%%%%%%%%
\subsection{Theoretical framework and Setup}
We consider the EMRIs described by the action \cite{Maselli:2020zgv}
\begin{equation}\label{eq:action}
S[g_{\mu\nu},\phi,\Psi]=S_0[g_{\mu\nu},\phi]+\alpha S_c[g_{\mu\nu},\phi]+S_m[g_{\mu\nu},\phi,\Psi],
\end{equation}
with
\begin{equation}
S_0=\int d^4x \frac{\sqrt{-g}}{16\pi}\left(R-\frac{1}{2}\partial_{\mu}\phi\partial^{\mu}\phi\right),
\end{equation}
where $R$ is the Ricci scalar.
$S_c$ describes the nonminimal coupling between the scalar field $\phi$ and the metric tensor, $\alpha$ is the coupling parameter with dimensions $[\alpha]=(mass)^n$.
The matter field action $S_m[g_{\mu\nu},\phi,\Psi]$ describes the spinning secondary.
% with the mass function $m(\phi)$.
Varying the action Eq.(\ref{eq:action}), one obtains the equations of motion
\begin{eqnarray}\label{eq:eom}
G_{\mu \nu}=R_{\mu \nu}-\frac{1}{2} g_{\mu \nu} R&=&T_{\mu \nu}^{\text{scal}}+\alpha T_{\mu \nu}^{c}+T_{\mu \nu}^{p},\label{eq:eom1}\\
\square \phi+\frac{16 \pi \alpha}{\sqrt{-g}} \frac{\delta S_{c}}{\delta \phi}&=&16 \pi\ \mathcal{T}_{scalar} ,\label{eq:eom2}
\end{eqnarray}
where $T_{\mu \nu}^{\text{scal}}=\frac{1}{2} \partial_{\mu} \phi \partial_{\nu} \phi-\frac{1}{4} g_{\mu \nu}(\partial \phi)^{2}$ is the stress-energy tensor of the scalar field, $\alpha T_{\mu \nu}^{c}$ is the stress-energy of the coupling term, and $T_{\mu\nu}^p$ represents the stress-energy tensor of the spinning secondary.
$\mathcal{T}_{scalar}$ is the source term which is obtained by varying $S_m$ with respect to $\phi$.

%while the stress-energy tensor of the spinning secondary is
%\begin{equation}\label{eq:sourceT}
%T_p^{\mu \nu}=m(\phi) \int \mathrm{d} \lambda\left[\frac{\delta^{(4)}\left(x-y_{p}(\lambda)\right)}{\sqrt{-g}} u^{(\mu} v^{\nu)}-\nabla_{\sigma}\left(S^{\sigma(\mu} v^{\nu)} \frac{\delta^{(4)}\left(x-y_{p}(\lambda)\right)}{\sqrt{-g}}\right)\right].
%\end{equation}
%Where $u^\mu$ is the normalized momenta and $S^{\mu\nu}$ is a skew-symmetric tensor.

%Based on the discussion in \cite{Maselli:2020zgv,Guo:2022euk}, the simplification is obtained by neglecting the higher-order infinitesimal terms of the equations of motion. 
By using the skeletonization approximation, the scalar field can be approximated to $\phi=\phi_0+\frac{m_p d}{r}+...$ far away from the matter source, where $\phi_0$ represents the background value of the scalar field, $d$ denotes the dimensionless scalar charge of the test body, and $m_p$ is the mass of the secondary.
Noticed that the scalar field in spacetime is directly coupled to the geometry, and the interaction between the scalar field and the secondary is reflected by a mass function $m(\phi)$.
After simplification, it is convenient to obtain the relation $m(\phi_0)=m_p$ and $m^{\prime}(\phi_0)/m(\phi_0)=-d/4$.

Therefore, the gravitational perturbation is described by a spinning secondary with mass $m(\phi_0)=m_p$ spiraling into a supermassive Kerr BH. 
Following the discussion in \cite{Piovano:2020zin}, the stress-energy tensor $T_{\mu\nu}^p$ reduces to the stress-energy tensor of a spinning test particle
\begin{equation}\label{eq:sourceT}
	T_{\mu\nu}^p=8\pi\,m_{p}\int \mathrm{d} \lambda\left[\frac{\delta^{(4)}\left(x-y_{p}(\lambda)\right)}{\sqrt{-g}} u^{(\mu} v^{\nu)}-\nabla_{\sigma}\left(S^{\sigma(\mu} v^{\nu)} \frac{\delta^{(4)}\left(x-y_{p}(\lambda)\right)}{\sqrt{-g}}\right)\right].
\end{equation}
As we neglect the directive interaction between the scalar field and the secondary spin which is higher-order infinitesimal interaction in this model, the scalar perturbation is sourced by the trajectory motion of the spinning secondary. 
Consequently with the result mentioned in \cite{Maselli:2020zgv,Guo:2022euk}, the source term of the scalar field reduces to
\begin{equation}
	\mathcal{T}_{scalar}=-\frac{d}{4}m_p\int\frac{dt}{v^t} \frac{\delta^{(4)}\left(x-y_{p}(t)\right)}{\sqrt{-g}}.
\end{equation}
As a result, the equations of motion can be expressed as
\begin{align}
G_{\mu\nu}&=T^{\rm p}_{\mu\nu}=8\pi\, m_{p}\int \mathrm{d} \lambda\left[\frac{\delta^{(4)}\left(x-y_{p}(\lambda)\right)}{\sqrt{-g}} u^{(\mu} v^{\nu)}-\nabla_{\sigma}\left(S^{\sigma(\mu} v^{\nu)} \frac{\delta^{(4)}\left(x-y_{p}(\lambda)\right)}{\sqrt{-g}}\right)\right]\label{eq:pertG},\\
  \square\phi&=-4\pi d\,m_{\rm p}\int\frac{dt}{v^t} \frac{\delta^{(4)}(x-y_{\rm p}(t))}{\sqrt{-g}}\,.\label{eq:pertphi}
\end{align}
 where $y_p$ is the worldline of the secondary, and $\lambda$ is the affine parameter which is set as the proper time. 
 The 4-velocity and normalized momenta of the secondary are represented by $v^\mu$ and $u^\mu$, respectively. Additionally, we use a skew-symmetric tensor $S^{\mu\nu}$ to derive the spin parameter, which can be obtained using $S^{2} \equiv \frac{1}{2} S^{\mu \nu} S_{\mu \nu}$. 
 To simplify our discussion below, we introduce the reduced spin parameter $\chi=\sigma/q$ where $\sigma=S/(m_p M)$ is the related dimensionless spin parameter and the mass ratio is $q=m_p/M$. 
For a more detailed discussion of the spinning orbital evolution please refer to subsection \ref{sec=motion}.

In this paper, we study a spinning secondary adiabatically spiraling into the central supermassive Kerr BH with the quasi-circular orbital evolution in the equatorial plane. 
The background Kerr metric reads
\begin{equation}
d s^{2}=-\left(1-\frac{2 M r}{\Sigma}\right) d t^{2}+\frac{\Sigma}{\Delta} d r^{2}-\frac{4 M a r \sin ^{2} \theta}{\Sigma} d t d \varphi+\Sigma d \theta^{2}+\frac{\sin ^{2} \theta}{\Sigma}\left(\varpi^{4}-a^{2} \Delta \sin ^{2} \theta\right) d \varphi^{2},
\end{equation}
with $\Sigma \equiv r^{2}+a^{2} \cos ^{2} \theta, \Delta \equiv r^{2}-2 M r+a^{2}, \varpi \equiv \sqrt{r^{2}+a^{2}}$, $M$ and $a$ is the mass and the spin of the Kerr black hole, respectively. 
Here we introduce the dimensionless parameters as $\hat{a}=a/M,\hat{t}=t/M,\hat{r}=r/M$, so we have $\{\hat{\Sigma},\hat{\Delta}, \hat{\varpi}\}=\{\Sigma, \Delta, \varpi\}/M^2$. 
The inner and outer horizon is given by $\hat{r}_{\pm}=1\pm\sqrt{1-\hat{a}^2}$, and the tortoise coordinate is defined by $d\hat{r}/\hat{r}_*=\hat{\Delta}/(\hat{r}^2+\hat{a}^2)$. 

%%%%%%%%%%%%%%%%%%%%%%%%%%%%%%%%%%%%%%%%%%%%%%%%%%%%%%%%%%%%%%%%%%%%%%%%%%%%
\subsection{Orbital motion of spin particle}\label{sec=motion}

In this EMRI system, the size of the secondary is significantly smaller than that of the central BH, so its stress-energy tensor $T_{\mu\nu}$ can be approximated by a multipolar expansion within gravitational skeletonization. 
Treating the secondary as a spinning particle corresponds to retaining only the first two multipoles. 
The covariant conservation of the energy-momentum tensor results in the Mathisson- Papapetrou- Dixon (MPD) equations \cite{Piovano:2020zin}
\begin{equation}
\begin{aligned}
\frac{\mathrm{d} y_p^{\mu}}{\mathrm{d} \lambda} &=v^{\mu} \\
\nabla_{\vec{v}} p^{\mu} &=-\frac{1}{2} R_{\nu \alpha \beta}^{\mu} v^{\nu} S^{\alpha \beta} \\
\nabla_{\vec{v}} S^{\mu \nu} &=2 p^{[\mu} v^{\nu]} \\
\mathfrak{m} & \equiv-p_{\mu} v^{\mu}
\end{aligned}
\end{equation}
where the 4-velocity $v^\mu$ is defined by the worldline $y_p^\mu(\lambda)$, and $\nabla_{\vec{v}} \equiv v^{\mu} \nabla_{\mu}$. 
The spin parameter $S$ is defined by the skew-symmetric tensor as $S^{2} \equiv \frac{1}{2} S^{\mu \nu} S_{\mu \nu}$, the linear momentum and 4-velocity are not aligned with $p^{\mu}=\frac{1}{v^{2}}\left(\mathfrak{m} v^{\mu}-v_{\sigma} \nabla_{\vec{v}} S^{\mu \sigma}\right)$ since $\mathfrak{m}$ represents the monopole rest-mass. 
Here we introduce the dynamical rest mass of the point particle as $m_p^2=-p^\sigma p_\sigma$. So the normalized momenta is given by $u^\mu=p^\mu/m_p$ which satisfies $u^\mu u_\mu=-1$.

%Following \cite{Piovano:2020zin}, 
We take the spin-supplementary condition by the Tulczyjew-Dixon equation
\begin{equation}
S^{\mu \nu} p_{\nu}=0.
\end{equation}
By incorporating the Kerr metric into these equations, the MPD equations can be expressed in exact formulas in Boyer-Lindquist coordinates
\begin{align}\label{eq:geomo1}
&\Sigma_{\sigma} \Lambda_{\sigma} \frac{\mathrm{d} \hat{t}}{\mathrm{~d} \hat{\lambda}}=\hat{a}\left(1+\frac{3 \sigma^{2}}{\hat{r} \Sigma_{\sigma}}\right)\left[\hat{J}_{z}-\hat{E}(\hat{a}+\sigma)\right]+\frac{\hat{r}^{2}+\hat{a}^{2}}{\hat{\Delta}} P_{\sigma}\\
&\left(\Sigma_{\sigma} \Lambda_{\sigma}\right)^{2}\left(\frac{\mathrm{d} \hat{r}}{\mathrm{~d} \hat{\lambda}}\right)^{2}=R_{\sigma}^{2} \\
&\Sigma_{\sigma} \Lambda_{\sigma} \frac{\mathrm{d} \phi}{\mathrm{d} \hat{\lambda}}=\left(1+\frac{3 \sigma^{2}}{\hat{r} \Sigma_{\sigma}}\right)\left[\hat{J}_{z}-\hat{E}(\hat{a}+\sigma)\right]+\frac{\hat{a}}{\hat{\Delta}} P_{\sigma}\label{eq:geomo3}
\end{align}
with
\begin{equation}
\begin{aligned}
&\Lambda_{\sigma}=1-\frac{3 \sigma^{2} \hat{r}\left[-(\hat{a}+\sigma) \hat{E}+\hat{J}_{z}\right]^{2}}{\Sigma_{\sigma}^{3}} \\
&R_{\sigma}=P_{\sigma}^{2}-\hat{\Delta}\left(\frac{\Sigma_{\sigma}^{2}}{\hat{r}^{2}}+\left[-(\hat{a}+\sigma) \hat{E}+\hat{J}_{z}\right]^{2}\right) \\
&P_{\sigma}=\left[\left(\hat{r}^{2}+\hat{a}^{2}\right)+\frac{\hat{a} \sigma}{\hat{r}}(\hat{r}+1)\right] \hat{E}-\left[\hat{a}+\frac{\sigma}{\hat{r}}\right] \hat{J}_{z},
\end{aligned}
\end{equation}
here $\Sigma_{\sigma}=\hat{r}^{2}\left(1-\frac{\sigma^{2}}{\hat{r}^{3}}\right)>0$. It is convenient to demonstrate that the geodesic equations given by Eq.\eqref{eq:geomo1}-\eqref{eq:geomo3} reduce to nonspinning geodesic motion when $\sigma\rightarrow 0$. 
In this EMRI model, we focus on the circular orbital motion on the equatorial plane which implies that both the radial velocity and radial acceleration are zero. 
The details of the calculation are omitted here but it can be found in \cite{Piovano:2020zin}. After simplification, we can determine the orbital frequency as measured by a static observer located at infinity 
\begin{equation}\label{eq:motionfre}
\hat{\Omega}=M\Omega=\frac{(2 \hat{a}+3 \sigma) \hat{r}^{3}+3\left(2 \hat{a}^{2} \sigma+\hat{a} \sigma^{2}\right) \hat{r}+4 \hat{a} \sigma^{2} \mp \hat{r} \mathcal{D}}{2\left(\hat{a}^{2}+3 \hat{a} \sigma+\sigma^{2}\right) \hat{r}^{3}+6 \sigma(\hat{a}+\sigma) \hat{a}^{2} \hat{r}+4 \hat{a}^{2} \sigma^{2}-2 \hat{r}^{6}},
\end{equation}
where 
\begin{equation}
\mathcal{D}=\sqrt{4 \hat{r}^{7}+12 \hat{a} \sigma \hat{r}^{5}+13 \sigma^{2} \hat{r}^{4}+6 \hat{a} \sigma^{3} \hat{r}^{2}-8 \sigma^{4} \hat{r}+9 \hat{a}^{2} \sigma^{4}}.
\end{equation}
The first integrals, of the spinning particle's motion, which are the orbital energy $\hat{E}$ and the orbital angular momentum $\hat{J}_z$ can be expressed by
\begin{eqnarray}
	\hat{E} &=&\frac{E}{M}=\frac{\hat{r} \sqrt{\hat{\Delta}}+(\hat{a} \hat{r}+\sigma) U_{\mp}}{\hat{r}^{2} \sqrt{1-U_{\mp}^{2}}}\label{eq:motionE}, \\
	\hat{J}_{z} &=&\frac{J_z}{m_pM}=\frac{\hat{r} \sqrt{\hat{\Delta}}(\hat{a}+\sigma)+\left[\hat{r}^{3}+\hat{r} \hat{a}(\hat{a}+\sigma)+\hat{a} \sigma\right] U_{\mp}}{\hat{r}^{2} \sqrt{1-U_{\mp}^{2}}}\label{eq:motionJ},
\end{eqnarray}
with
\begin{equation}
U_{\mp}=-\frac{2 \hat{a} \hat{r}^{3}+3 \sigma \hat{r}^{2}+\hat{a} \sigma^{2} \mp \mathcal{D}}{2 \sqrt{\hat{\Delta}}\left(\hat{r}^{3}+2 \sigma^{2}\right)},
\end{equation}
the sign $\mp$ represents prograde and retrograde orbits, respectively. 
The expressions of the conserved quantities \eqref{eq:motionE} and \eqref{eq:motionJ}, as well as the orbital frequency \eqref{eq:motionfre} will be useful when studying the adiabatic evolution of the spinning orbital motion.

%%%%%%%%%%%%%%%%%%%%%%%%%%%%%%%%%%%%%%%%%%%%%%%%%%%%%%%%%%%%%%%%%%%%%%%%%%%%
\section{Perturbations and data processing approach}\label{sec=perturb}
\subsection{Gravitational and scalar perturbation}

The wave equation of the metric perturbation can be obtained by the Teukolsky formalism, which is governed by the $\psi_4$ Weyl scalar:
\begin{equation}
\psi_{4}=\rho^{4} \sum_{\ell=2}^{\infty} \sum_{m=-\ell}^{\ell} \int_{-\infty}^{\infty} d \hat{\omega} R_{\ell m \hat{\omega}}(\hat{r})_{-2} S_{\ell m}^{\hat{a} \hat{\omega}}(\theta) e^{i(m \varphi-\hat{\omega} \hat{t})},
\end{equation}
where $\rho=(\hat{r}-i \hat{a}\cos\theta)^{-1}$, the $s=-2$ spin weighted orthonormal spheroidal harmonics is $_{-2} S_{\ell m}^{\hat{a} \hat{\omega}}$ with its eigenvalue $\lambda_G$. 
At infinity, the GW polarizations are given by the relation
\begin{equation}
\psi_4=\frac{1}{2}\frac{\partial^2}{\partial \hat{t}^2}(h_+-i h_{\times}).
\end{equation}
The radial Teukolsky equation is
\begin{equation}\label{eq:radialGR}
\hat{\Delta}^{2} \frac{d}{d \hat{r}}\left(\frac{1}{\hat{\Delta}} \frac{d R_{\ell m \omega}(\hat{r})}{d \hat{r}}\right)-V_G(\hat{r}) R_{\ell m \hat{\omega}}(\hat{r})=\mathcal{T}^T_{\ell m \hat{\omega}},
\end{equation}
with its effective potential
\begin{eqnarray}
	V_G(\hat{r}) &=&-\frac{K^{2}+4 i(\hat{r}-1) K}{\hat{\Delta}}+8 i \hat{\omega} \hat{r}+\lambda_G,
\end{eqnarray}
where $K=\left(\hat{r}^{2}+\hat{a}^{2}\right) \hat{\omega}-\hat{a} m$. 
The homogeneous Teukolsky equation has two linearly independent solutions which satisfy pure ingoing boundary conditions near horizon $R_{\ell m\hat{\omega}}^{in}$ and pure outgoing boundary conditions at infinity $R_{\ell m\hat{\omega}}^{out}$. 
Through the Green's function method, the solution of the inhomogeneous Teukolsky equation is obtained by
\begin{equation}
R_{\ell m \hat{\omega}}(\hat{r}) =\frac{1}{W_{G }}\left\{R_{\ell m \hat{\omega}}^{\mathrm{out}}(\hat{r}) \int_{\hat{r}_{+}}^{\hat{r}} \mathrm{~d} \hat{r}^{\prime} \frac{R_{\ell m \hat{\omega}}^{\mathrm{in}}\left(\hat{r}^{\prime}\right) \mathcal{T}^{T}_{\ell m \hat{\omega}}\left(\hat{r}^{\prime}\right)}{\hat{\Delta}^{2}}+R_{\ell m \hat{\omega}}^{\mathrm{in}}(\hat{r}) \int_{\hat{r}}^{\infty} \mathrm{d} \hat{r}^{\prime} \frac{R_{\ell m \hat{\omega}}^{\mathrm{out}}\left(\hat{r}^{\prime}\right) \mathcal{T}^{T}_{\ell m \hat{\omega}}\left(\hat{r}^{\prime}\right)}{\hat{\Delta}^{2}}\right\},
\end{equation}
with the constant Wronskian $W_G\equiv R^{\rm in}_{\ell m\hat{\omega}}dR^{\rm out}_{\ell m\hat{\omega}}/d\hat{r}_*-R^{\rm out}_{\ell m\hat{\omega}}dR^{\rm in}_{\ell m\hat{\omega}}/d\hat{r}_*$. 
The source term $\mathcal{T}^T_{\ell m \hat{\omega}}$ is obtained by the stress-energy tensor in Eq.\eqref{eq:pertG} by the Newman-Penrose formalism, more detail can be referred by \cite{Chrzanowski:1976jy,Piovano:2020zin}.

The solution of the inhomogeneous Teukolsky equation also satisfies pure ingoing boundary conditions near the horizon and pure outgoing boundary conditions at infinity
\begin{equation}
\begin{aligned}
R_{\ell m \hat{\omega}}\left(\hat{r} \rightarrow \hat{r}_{+}\right) &=Z_{\ell m \hat{\omega}}^{\infty} \hat{\Delta}^{2} e^{-i \hat{\kappa} \hat{r}_{*}}, \\
R_{\ell m \hat{\omega}}(\hat{r} \rightarrow \infty) &=Z_{\ell m \hat{\omega}}^{H} \hat{r}^{3} e^{i \hat{\omega} \hat{r}_{*}},
\end{aligned}
\end{equation}
with the coefficients
\begin{equation}
\begin{aligned}
&Z_{\ell m \hat{\omega}}^{\infty}=C_{\ell m \hat{\omega}}^{\infty} \int_{\hat{r}_{+}}^{\infty} \mathrm{d} \hat{r}^{\prime} \frac{R_{\ell m \hat{\omega}}^{\mathrm{out}}\left(\hat{r}^{\prime}\right)}{\hat{\Delta}^{2}} \mathcal{T}^{T}_{\ell m \hat{\omega}}\left(\hat{r}^{\prime}\right), \\
&Z_{\ell m \hat{\omega}}^{H}=C_{\ell m \hat{\omega}}^{H} \int_{\hat{r}_{+}}^{\infty} \mathrm{d} \hat{r}^{\prime} \frac{R_{\ell m \hat{\omega}}^{\mathrm{in}}\left(\hat{r}^{\prime}\right)}{\hat{\Delta}^{2}} \mathcal{T}^{T}_{\ell m \hat{\omega}}\left(\hat{r}^{\prime}\right),
\end{aligned}
\end{equation}
where $C_{\ell m \hat{\omega}}^{H,\infty}$ can be found by Eq.(86) in \cite{Piovano:2020zin}. 
For simplicity, $\varphi(\hat{t})=\hat{\Omega} \hat{t}$ with a equatorial circular orbit, and we have
\begin{equation}
Z_{\ell m \hat{\omega}}^{H, \infty}=\delta(\hat{\omega}-m \hat{\Omega}) \mathcal{A}_{\ell m \hat{\omega}}^{H, \infty}.
\end{equation}

The scalar perturbation is expanded by the scalar spheroidal harmonics
\begin{equation}
\phi(\hat{t}, \hat{r}, \theta, \varphi)=\sum_{\ell, m} \int d\hat{\omega}~e^{i(m \varphi-\hat{\omega}\hat{t})}\frac{X_{\ell m\hat{\omega}}(r)}{\sqrt{\hat{r}^{2}+\hat{a}^{2}}} {}_0 S_{\ell m}(\theta),
\end{equation}
the $s=0$ orthonormal spheroidal harmonics is ${}_0 S_{\ell m}(\theta)$ with the eigenvalue $\lambda_s$.
We write the radial scalar perturbation equation here
\begin{equation}\label{eq:radialsca}
\left[\frac{d^{2}}{d \hat{r}_{*}^{2}}+V_{s}(\hat{r})\right] X_{\ell m \hat{\omega}}(\hat{r})=\frac{\hat{\Delta}}{\left(\hat{r}^{2}+\hat{a}^{2}\right)^{3 / 2}} \mathcal{T}^s_{\ell m \hat{\omega}},
\end{equation}
with its effective potential
\begin{equation}
V_{s}=\left(\hat{\omega}-\frac{\hat{a} m}{\vartheta}\right)^{2}-\frac{\hat{\Delta}}{\vartheta^{4}}\left[\lambda_s ~\vartheta^{2}+2 \hat{r}^{3}+\hat{a}^{2}\left(\hat{r}^{2}-4 \hat{r}+\hat{a}^{2}\right)\right],
\end{equation}
where $\vartheta=\hat{r}^2+\hat{a}^2$, and $\mathcal{T}^s_{\ell m\hat{\omega}}$ is constructed from the source term on the right-hand side of Eq. \eqref{eq:pertphi}.
Similar to the case of gravitational perturbations, the homogeneous scalar perturbation equation has two linearly independent solutions, namely $X_{\ell m\hat{\omega}}^{in,out}$, which satisfy pure ingoing boundary conditions near the horizon and pure outgoing boundary conditions at infinity respectively. 
Using the Green's function method, the solution of the inhomogeneous equation can be constructed by
\begin{equation}
X_{\ell m \hat{\omega}}(\hat{r}) =X_{\ell m \hat{\omega}}^{\rm out}(\hat{r}) \int_{\hat{r}_{+}}^{\hat{r}} ~d r^{\prime} \frac{X_{\ell m \hat{\omega}}^{\rm in}\left(r^{\prime}\right) \mathcal{T}^s_{\ell m \hat{\omega}}\left(r^{\prime}\right)}{W_{s}}+X_{\ell m \hat{\omega}}^{\rm in}(\hat{r}) \int_{\hat{r}}^{\infty} d r^{\prime} \frac{X_{\ell m \hat{\omega}}^{\rm out}\left(r^{\prime}\right) \mathcal{T}^s_{\ell m \hat{\omega}}\left(r^{\prime}\right)}{W_{s}},
\end{equation}
Likewise, the constant Wronskian is $W_s\equiv X^{\rm in}_{\ell m\hat{\omega}}dX^{\rm out}_{\ell m\hat{\omega}}/d\hat{r}_*-X^{\rm out}_{\ell m\hat{\omega}}dX^{\rm in}_{\ell m\hat{\omega}}/d\hat{r}_*$.
And the inhomogeneous solution of the scalar perturbation can also give the boundary condition so that
\begin{align}
	&X_{\ell m \hat{\omega}}\left(\hat{r} \rightarrow \hat{r}_{+}\right) =\mathcal{Z}_{\ell m \hat{\omega}}^{\infty} e^{-i \hat{\kappa} \hat{r}_{*}}, \\
    &X_{\ell m \hat{\omega}}(\hat{r} \rightarrow \infty) =\mathcal{Z}_{\ell m \hat{\omega}}^{H} e^{i \hat{\omega} \hat{r}_{*}},
\end{align}
with the coefficients
\begin{equation}
\mathcal{Z}_{\ell m \hat{\omega}}^{H, \infty}=-4\pi dq\frac{X_{\ell m \hat{\omega}}^{\rm in,up}\left(\hat{r}_{\rm p}\right)}{W_s\ u^{t}}\ \frac{{}_0 S_{\ell m}^{*}(\pi / 2)}{\sqrt{\hat{r}_{\rm p}^{2}+\hat{a}^{2}}},
\end{equation}
where ${}_0S_{\ell m}(\theta)^*$ is its complex conjugation and $\hat{\kappa}=m\hat{\Omega}-m\hat{a}/(2\hat{r}_+)$.

After we have gotten the solution for both the gravitational perturbation and scalar perturbation, next we can compute the energy fluxes of this model \cite{Teukolsky:1973ha,Yunes:2011aa}. 
From the gravitational perturbation part, the energy fluxes at the horizon and at infinity are
\begin{align}
	\dot{E}_{T}^{H}&=\sum_{\ell=2}^{\infty} \sum_{m=1}^{\ell} \alpha_{\ell m} \frac{\left|\mathcal{A}_{\ell m \hat{\omega}}^{\infty}\right|^{2}}{2\pi(m \hat{\Omega})^{2}},\\
	\dot{E}_{T}^{\infty}&=\sum_{\ell=2}^{\infty} \sum_{m=1}^{\ell} \frac{\left|\mathcal{A}_{\ell m \hat{\omega}}^{H}\right|^{2}}{2\pi(m \hat{\Omega})^{2}},
\end{align}
with the coefficients
\begin{align}
\alpha_{\ell m}&=\frac{256\left(2 \hat{r}_{+}\right)^{5} \hat{\kappa}\left(\hat{\kappa}^{2}+4 \epsilon^{2}\right)\left(\hat{\kappa}^{2}+16 \epsilon^{2}\right)(m \hat{\Omega})^{3}}{\left|C_{\ell m}\right|^{2}},\\
\left|C_{\ell m}\right|^{2}&=\left[\left(\lambda_G+2\right)^{2}+4 \hat{a}(m \hat{\Omega})-4 \hat{a}^{2}(m \hat{\Omega})^{2}\right]\times\left[\lambda_G^{2}+36 m \hat{a}(m\hat{\Omega})-36\hat{a}^{2}(m \hat{\Omega})^{2}\right] \\
&+\left(2 \lambda_G+3\right)\left[96 \hat{a}^{2}(m \hat{\Omega})^{2}-48 m \hat{a}(m \hat{\Omega})\right]+144(m \hat{\Omega})^{2}\left(1-\hat{a}^{2}\right)
\end{align}
where $\epsilon=\sqrt{1-\hat{a}^2}/(4\hat{r}_+)$.
From the scalar perturbation part, the energy fluxes are
\begin{align}
\dot{E}_{s}^{H}&=\frac{1}{16\pi}\sum_{\ell=1}^{\infty} \sum_{m=-\ell}^{\ell} m ~\hat{\Omega} ~\hat{\kappa}\left|\mathcal{Z}_{l m \omega}^{\infty}\right|^{2},\\
\dot{E}_{s}^{\infty}&=\frac{1}{16\pi}\sum_{\ell=1}^{\infty} \sum_{m=-\ell}^{\ell} m^2 ~\hat{\Omega}^2\left|\mathcal{Z}_{\ell m \omega}^{H}\right|^{2}.
\end{align}

 By utilizing the code from the Black Hole Perturbation Toolkit \cite{BHPToolkit, Piovano:2020zin}, we can numerically solve these perturbation equations. The total energy fluxes of the EMRI system are obtained from the numerical solutions
\begin{equation}
\mathcal{F}_{\rm tot}=\dot{E}_{T}+\delta\dot{E}_{s}=\dot{E}^H_T+\dot{E}^\infty_T+\delta\dot{E}^H_s+\delta\dot{E}^\infty_s,
\end{equation}
Where the subscript ``$T$" stands for tensor modes and subscript ``$s$" represents scalar modes, superscripts ``$H$" and ``$\infty$" refer to horizon and infinity, respectively. 
$\dot{E}_T$ and $\delta\dot{E}_s$ are the total gravitational energy flux and total scalar energy flux.

Using the calculated total energy fluxes, we can now determine the adiabatic evolution of the spinning secondary, which is balanced by the energy emissions
 \begin{equation}\label{eq:orbittime}
 \frac{d r}{dt}=-\mathcal{F}_{\rm tot}(t)\left(\frac{d E}{dr}\right)^{-1}\, , \frac{d\varphi}{dt}=\Omega(r(t)).
 \end{equation}
where orbital energy $E$ is given by Eq.\eqref{eq:motionE} and orbital frequency $\Omega$ is shown in Eq.\eqref{eq:motionfre}. 
Here $\varphi$ represents the orbital phase, and for the dominant mode, the GW phase is given by $N_{\chi}^d=\varphi_{GW}(t_{\rm end})=2\varphi(t_{\rm end})$, where $t_{\rm end}$ denotes the exit time when the evolution ends.

%%%%%%%%%%%%%%%%%%%%%%%%%%%%%%%%%%%%%%%%%%%%%%%%%%%%%%%%%%%%%%%%%%%%%%%%%%%%
\subsection{Data processing approach}

In the following numerical calculations, we set the parameters as follows: without loss of generality, the primary spin is $a=0.9M$, primary mass is $M=4\times 10^5M_{\odot}$, and the secondary mass is $m_p=10M_{\odot}$ so the mass ratio is $q=2.5\times 10^{-5}$. 
In the parameter space, we calculate the secondary spin in the range $\chi\in[0,0.5]$ and the scalar charge in the range $d\in[0,0.5]$. 
And summing all the multipole contributions up to $\ell=18$. 

Different from the original approach of simulating a one-year orbital evolution before plunging into ISCO, we adopt the modified approach commonly used in detecting of secondary spin \cite{Piovano:2020ooe,Piovano:2020zin,Rahman:2021eay,Skoupy:2022adh,Drummond:2022xej}.
In our approach, the spinning secondary starts at $r_{start}=11.53M$ and spirals inwards the central BH. After one year's evolution, the simulation is terminated near $r_{ISCO}$. 
Although our simulation ensures that the position of the secondary after one-year evolution is as close as possible to the ISCO, the accumulated phase obtained from our calculation will still be smaller than that of the original approach. 
However, we add an extra constraint on the initial position of the simulation, which greatly improves the results of the dephasing, enhancing the detection of GWs. 

For instance, we can contrast the amount of dephasing resulting from two distinct simulation methods. 
One approach involves a year-long orbital evolution followed by a plunge into the ISCO, while the other approach involves starting from the same initial position and undergoing a year-long evolution.
%As an example, let's compare the dephasing obtained by the two different simulating approaches, one with one-year orbital evolution before plunging into ISCO, and the other with one-year evolution starting at the same initial position. 
We define $N_{\chi}^d$ as the total GW phase of our model, so $N_{\chi=0}^d$ and $N^{d=0}_{\chi}$ are the GW phase caused by the scalar charge and secondary spin independently, and $N^{d=0}_{\chi=0}$ is the pure GR GW phase. 
In Appendix \ref{sec=data}, we present the data of dephasing $|N_{\chi}^d-N_{\chi=0}^d|$ which describes the dephasing caused by the secondary spin. 
TABLE \ref{tab:method1} shows the data simulated by our modified approach while TABLE \ref{tab:method2} displays the results obtained using the original approach.
The first column contains the data in GR, and the subsequent columns show the results in modified gravity with scalar charge $d$.
The modified approach exhibits significant advantages over the original approach.
It can obtain a larger dephasing than the original approach for each parameter. 
Even in the GR case, the modified approach improves the dephasing more greatly than the original approach. 
It also shows a larger increase in dephasing for each secondary spin $\chi$ with the increase of scalar charge $d$, indicating a significant improvement in the resolution and accuracy for the secondary spin $\chi$. More detail will be shown in the next section.

Additionally, it is necessary to indicate that the total GW phase of this model is not simply the summation of all contributions from the model parameters. 
The total phase summation is determined by $N_{\chi=0}^d+N^{d=0}_{\chi}-N^{d=0}_{\chi=0}$, while the third term removes the extra GR GW phase $N^{d=0}_{\chi=0}$ calculated in the first two terms. 
The GW phase difference between the total GW phase of this model and the GW phase summation is expressed by $|N_{\chi}^d-N_{\chi=0}^d-N^{d=0}_{\chi}+N^{d=0}_{\chi=0}|$, and the data is presented in TABLE \ref{tab:absphasing} of Appendix \ref{sec=data}. 
The results reveal that the GW phase difference resulting from the phase summation is more pronounced in the region with large values of $\chi$ and $d$ region.

%%%%%%%%%%%%%%%%%%%%%%%%%%%%%%%%%%%%%%%%%%%%%%%%%%%%%%%%%%%%%%%%%%%%%%%%%%%%
\section{Numerical results}\label{sec=result}

In this section, we will present the main findings of our modified gravity model and its implications for detecting the secondary spin in an EMRI system. Unlike pure GR, our model incorporates a scalar field to explain the effects of the scalar charge and the secondary spin on GW.
%%%%%%%%%%%%%%%%%%%%
\begin{figure}[thbp]
\center{
\includegraphics[width=3.in]{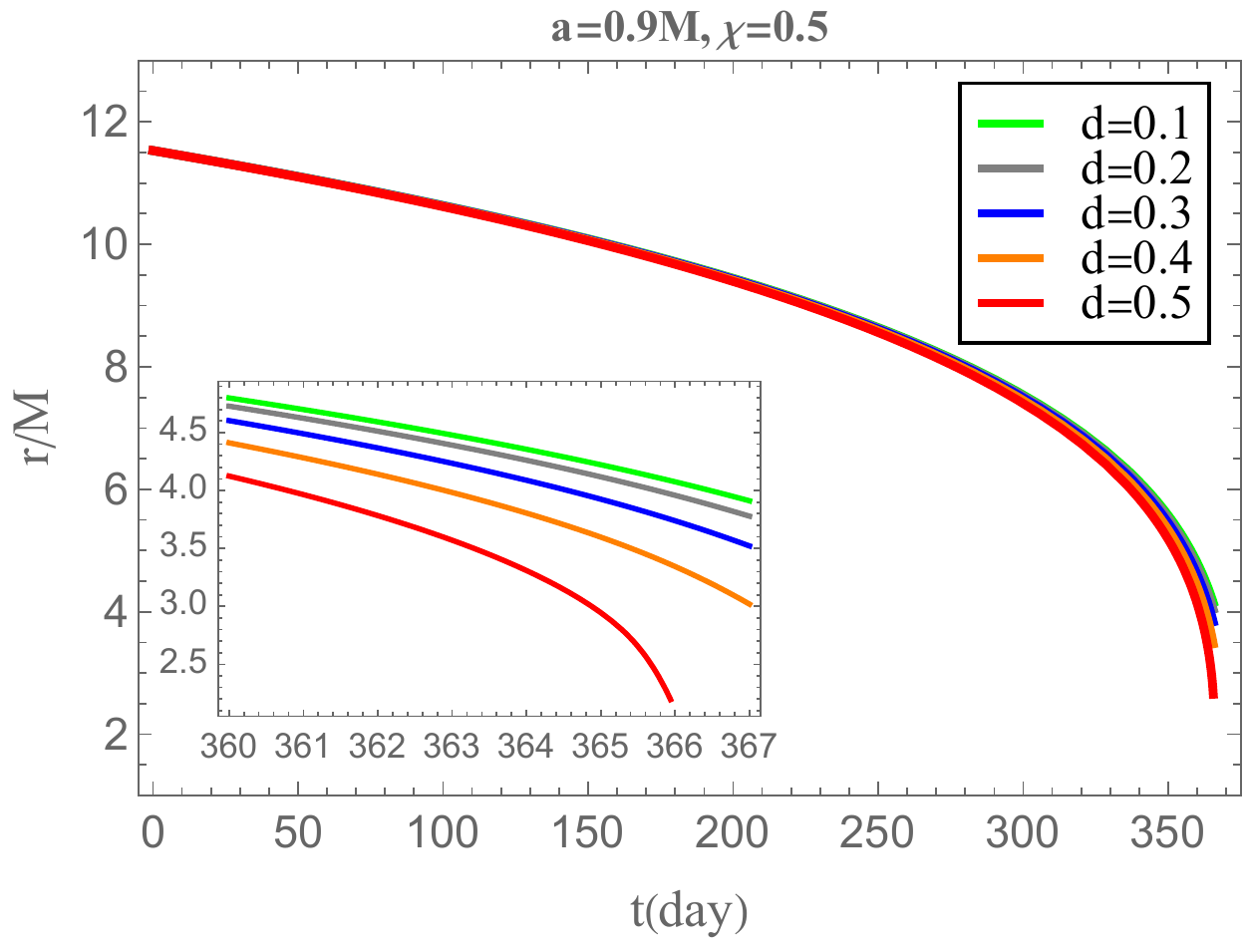}
\includegraphics[width=3.in]{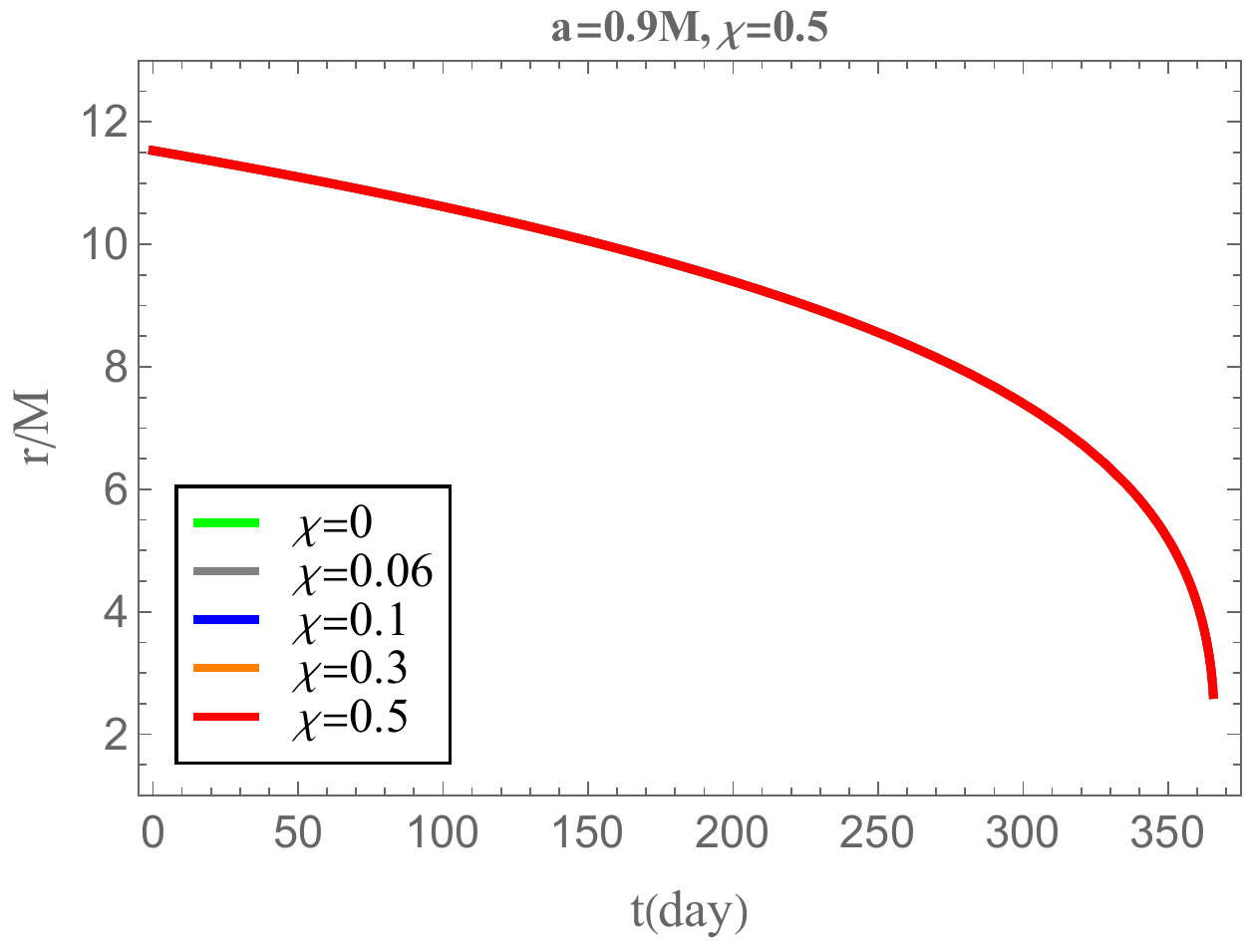}
\caption{Fixing $a=0.9M$, radial location of the secondary $r(t)$ as the function of the evolution time $t$. Left: the effect of different scalar charges on the orbital evolution when we set $\chi=0.5$; Right: the effect of different secondary spin $\chi$ on the orbital evolution when we set $d=0.5$.
}\label{fg:orbit}}
\end{figure}
%%%%%%%%%%%%%%%%%%%%

It is worth noting that the contribution of the scalar charge to the GW phase is comparable to that of GR phase with $\mathcal{O}(1/q)$ \cite{Maselli:2020zgv}, while the effect of the secondary spin is of higher-order with a factor of $\mathcal{O}(q^2)$ \cite{Burko:2015sqa}. 
This is demonstrated by the scalar perturbation equation \eqref{eq:pertphi}, where the scalar charge appears directly in the source term, leading to the scalar radiation proportional to $d^2$.
In contrast, the effect of the secondary spin on scalar radiation is only reflected in the 4-velocity in Eq. \eqref{eq:pertphi}. 

\begin{figure}[thbp]
\center{
\includegraphics[width=3in]{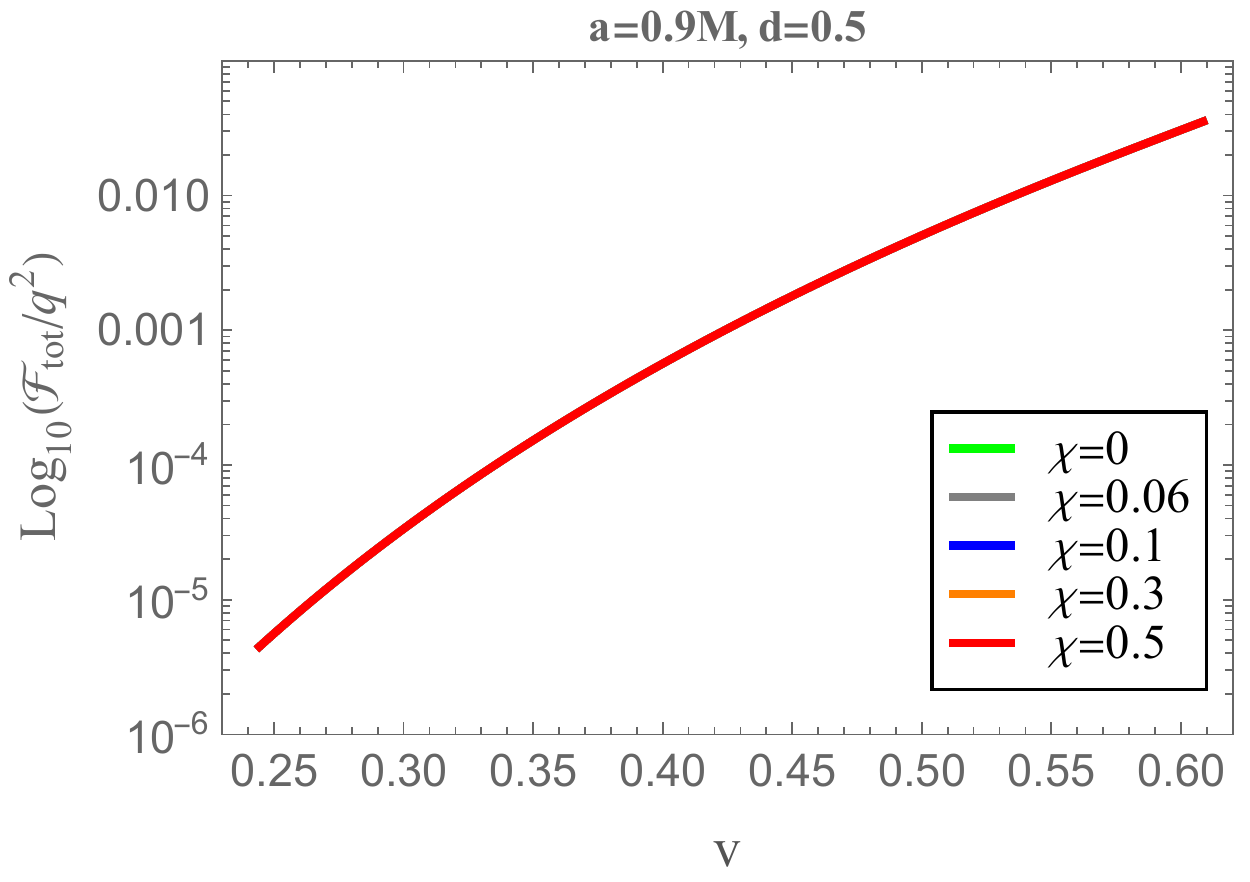}
\includegraphics[width=3in]{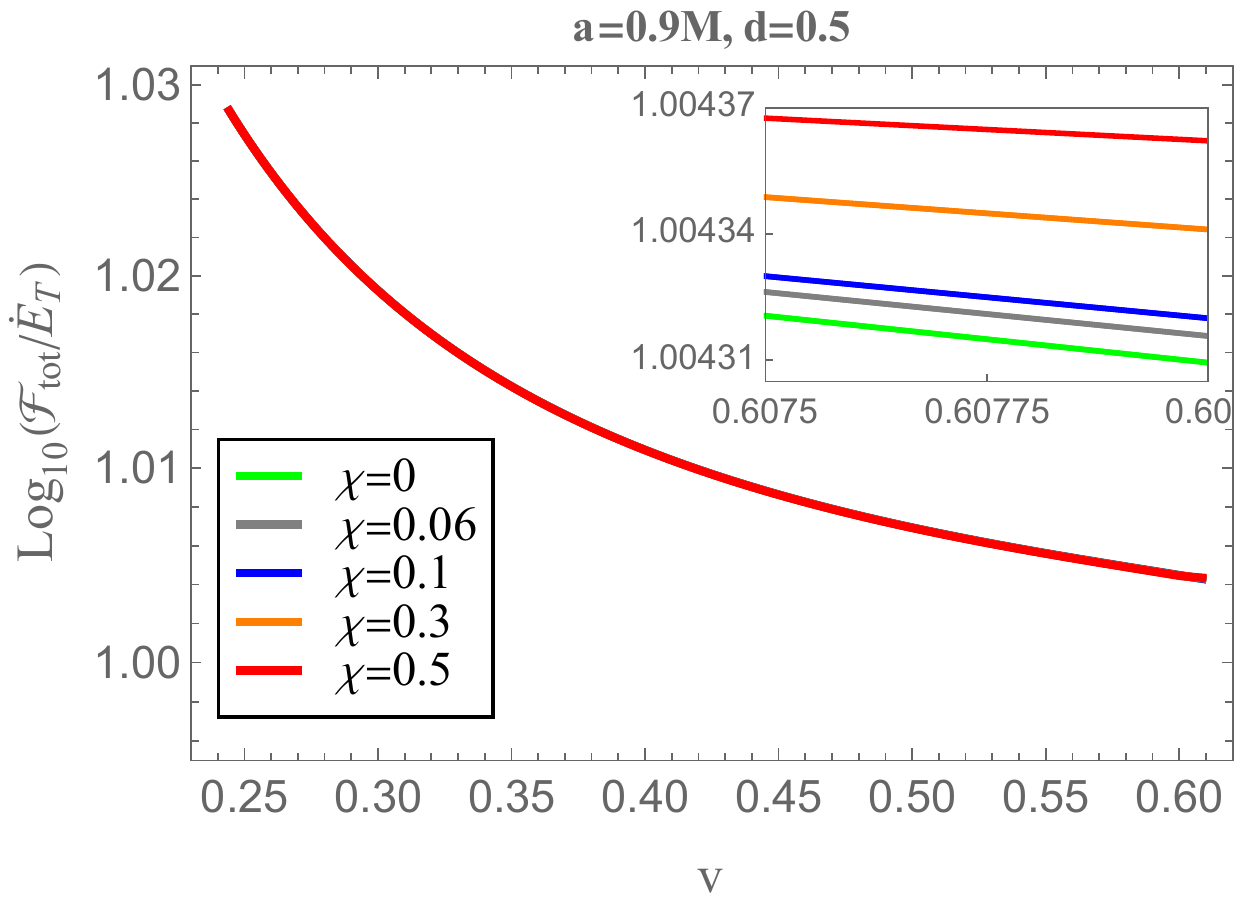}
\caption{Fixing $a=0.9M$, the total energy flux $\mathcal{F}_{tot}$ and the relative difference between total energy flux and gravitational energy flux $\mathcal{F}_{tot}/\dot{E}_T$ as a function of the orbital velocity $v=(M\Omega)^{1/3}$ with different secondary spin $\chi$ for scalar charge $d=0.5$.
}\label{fg:fluxd}}
\end{figure}

\begin{figure}[thbp]
\center{
\includegraphics[width=3in]{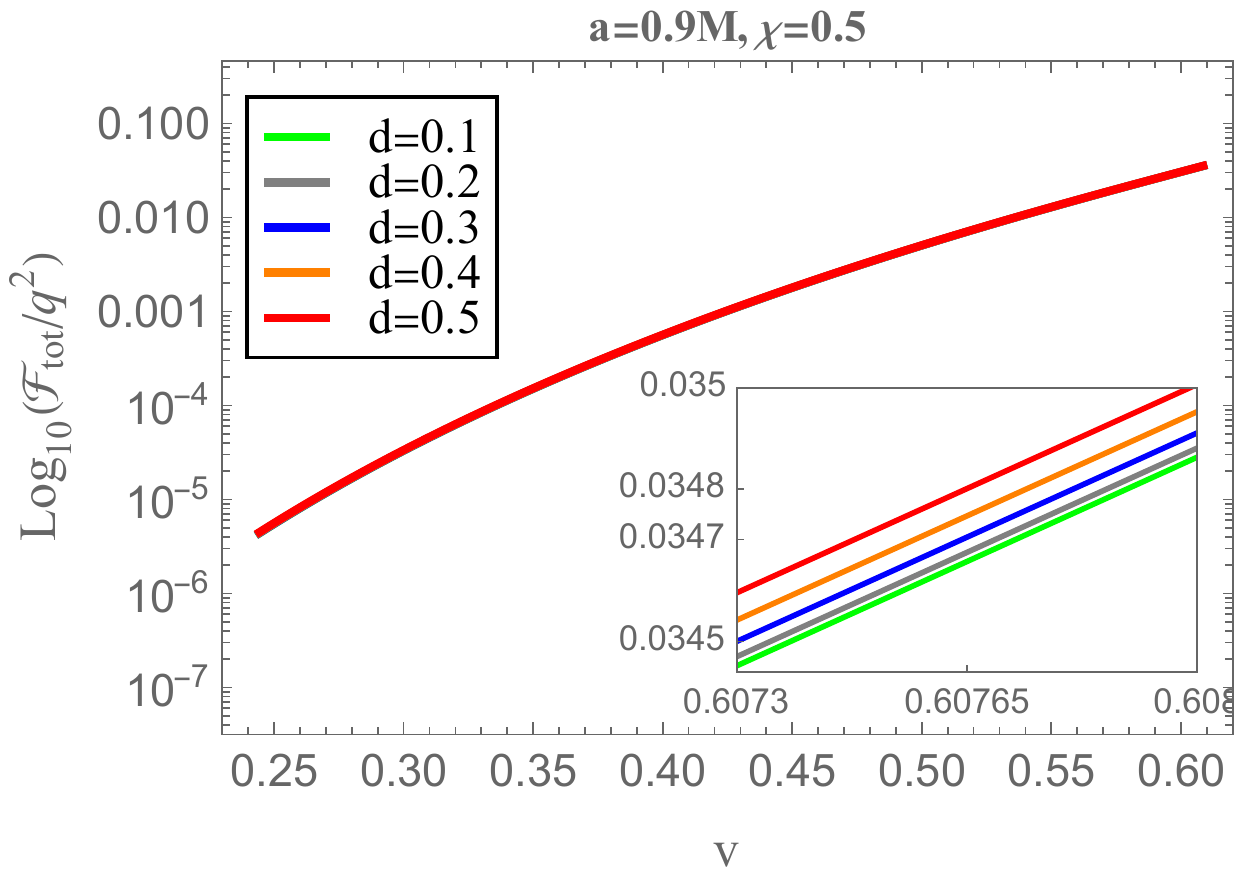}
\includegraphics[width=3in]{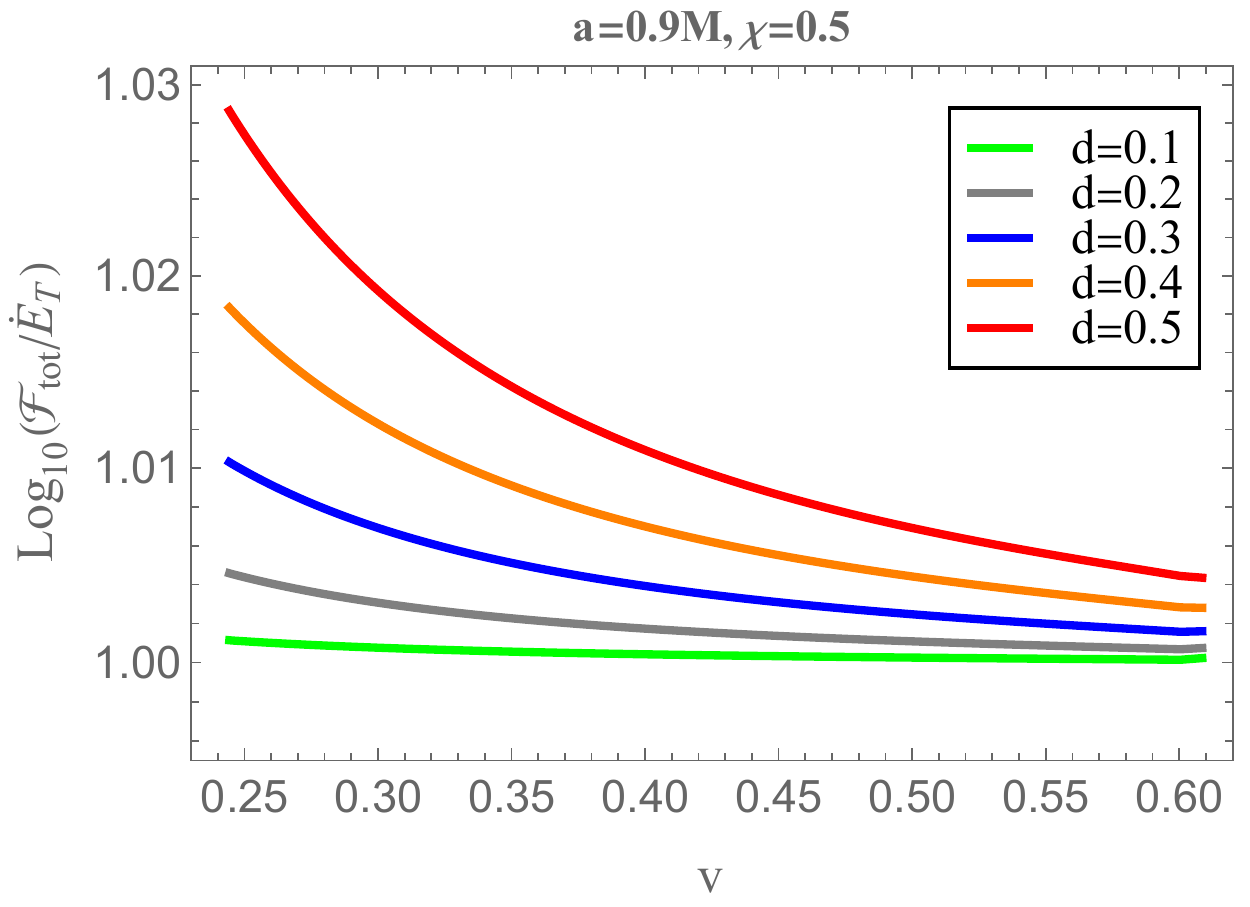}
\caption{Fixing $a=0.9M$, the total energy flux $\mathcal{F}_{tot}$ and the relative difference between total energy flux and gravitational energy flux $\mathcal{F}_{tot}/\dot{E}_T$ as a function of the orbital velocity $v=(M\Omega)^{1/3}$ with different secondary scalar charge $d$ for secondary spin $\chi=0.5$.
}\label{fg:fluxx}}
\end{figure}

The adiabatic evolution of the secondary, as shown in Fig. \ref{fg:orbit}, corroborates our discussion.
Notably, the presence of the scalar charge $d$ accelerates the fall of the spinning secondary into BH, while the changes in the secondary spin have a negligible impact on the orbital evolution. 
This conclusion also applies to the behavior of the energy fluxes, as illustrated in Fig. \ref{fg:fluxd} and Fig. \ref{fg:fluxx}. 
In Fig. \ref{fg:fluxd}, setting scalar charge $d=0.5$, an increasing secondary spin $\chi$ has little effect on the total energy flux $\mathcal{F}_{tot}$. 
This phenomenon becomes even more apparent when observing the ratio of total energy flux to GR energy flux $\mathcal{F}_{tot}/\dot{E}_T$ in the right figure. 
However, the presence of additional scalar radiation can amplify the orbital deviations and dephasing caused by the secondary spin, as evident in Fig. \ref{fg:fluxx}.
By fixing the secondary spin $\chi=0.5$, the growth of scalar charge leads to an enlargement of the total energy flux $\mathcal{F}_{tot}$, particularly noticeable in the ratio of total energy flux to GR $\mathcal{F}_{tot}/\dot{E}_T$. 

%%%%%%%%%%%%%%%%%%%%
%%%%%%%%%%%%%%%%%%%%

%%%%%%%%%%%%%%%%%%%%
%%%%%%%%%%%%%%%%%%%%

As we discussed earlier, the secondary spin in the EMRI system is a secondary effect that does not impact the detection of the scalar charge. 
Therefore, it is appropriate to overlook the influence of the secondary spin when designing the GW template for the detection of the scalar charge \cite{Piovano:2021iwv}. 
Interestingly, the existence of the scalar field would amplify both the deviation in the orbital evolution and the total energy radiation caused by the secondary spin, thereby, improving the ability to detect the secondary spin.

\begin{figure}[thbp]
\center{
\includegraphics[width=3in]{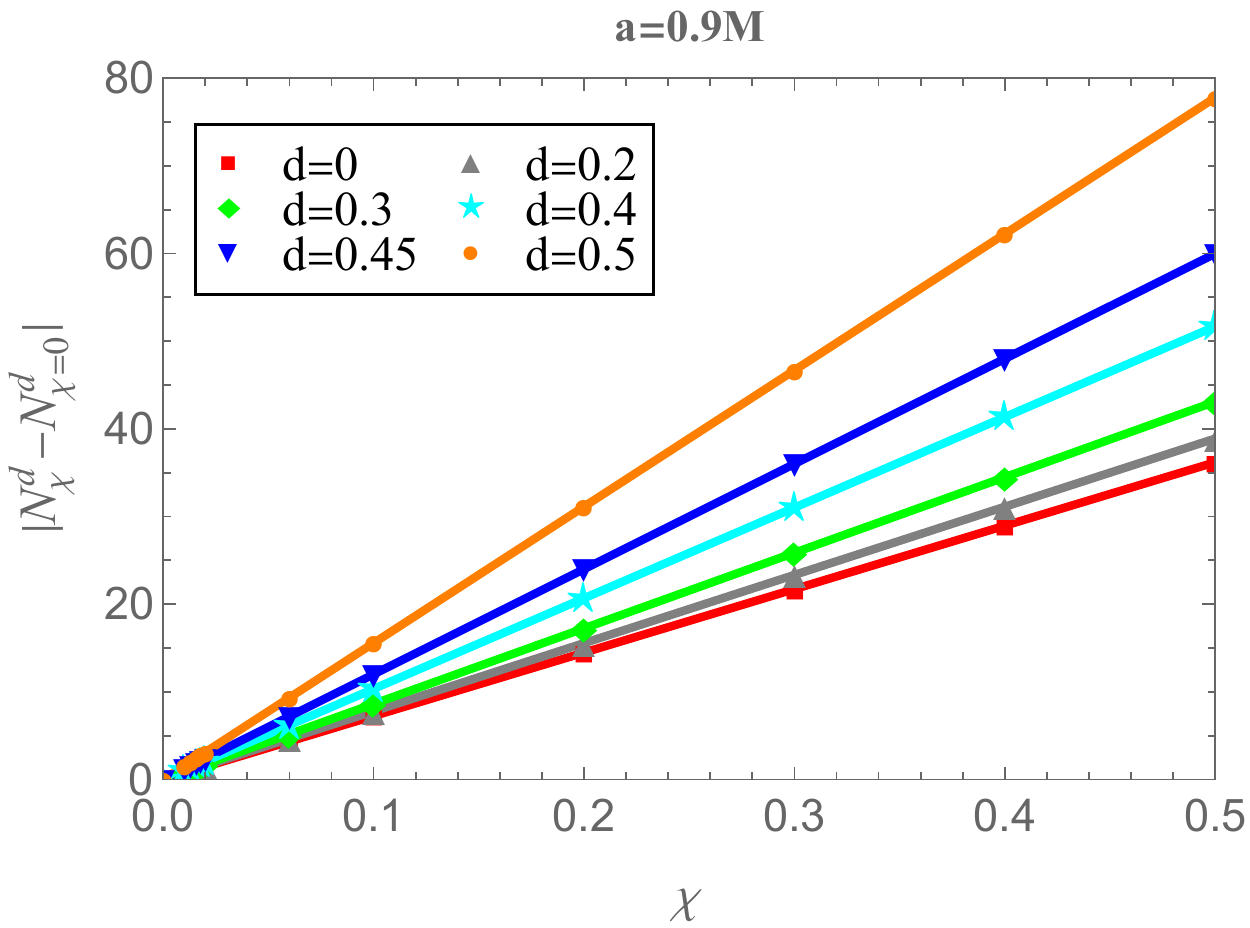}
\includegraphics[width=3in]{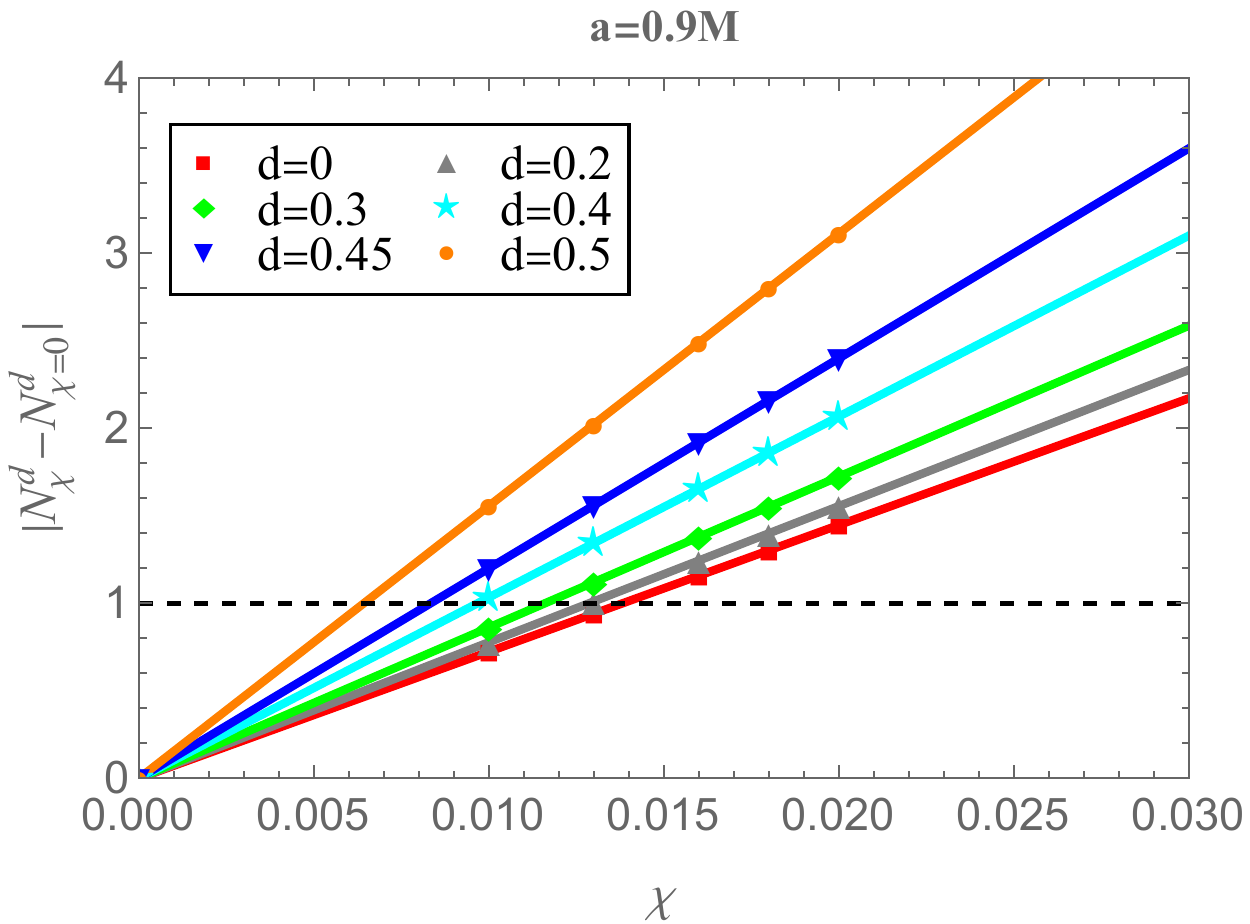}
\caption{Fixing $a=0.9M$, the behavior of dephasing $|N_{\chi}^d-N_{\chi=0}^d|$ as the function of secondary $\chi$. The right figure is a local enlargement of the left figure in the small $\chi$ area. The black horizontal dashed line represents the limit $\Delta\varphi= 1$ rad.}\label{fg:dephase}}
\end{figure}

This motivates us to further study the GW dephasing to facilitate the detection of the secondary spin in this model. 
By solving the equations of adiabatic evolution \eqref{eq:orbittime}, we obtain the total GW phase $N_{\chi}^d$ during the entire evolution. 
The dephasing is then calculated by $|N_{\chi}^d-N_{\chi=0}^d|$ as a function of the secondary spin $\chi$ for different scalar charge $d$. 
As shown in Fig. \ref{fg:dephase}, the red line represents the result in GR, where the dephasing linearly increases with the secondary spin, consistent with the discussion in \cite{Piovano:2020zin}. 
The presence of scalar charges significantly amplifies the dephasing in the model, particularly in regions where the secondary spin $\chi$ is relatively large. 
This suggests that the scalar charge $d$ can effectively improve the model's detection limit for the secondary spin $\chi$, as illustrated in the right figure. 
The phase resolution of a space-based GW detector is limited to $\Delta\varphi\lesssim 1$ rad by matched-filter search and parameter estimation \cite{Lindblom:2008cm}. 
Taking $\Delta\varphi= 1$ rad as a limit for discussion, when the scalar charge is $d=0$, the minimum detectable value for the secondary spin is $\chi=0.014$. However, when the scalar charge is increased to $d=0.5$, we can detect a spin of $\chi=0.006$, which is a $133\%$ improvement in the detection limit. 
The detailed data for dephasing can also be found in TABLE \ref{tab:method1} in Appendix \ref{sec=data}.

%%%%%%%%%%%%%%%%%%%%
%%%%%%%%%%%%%%%%%%%%

\begin{figure}[thbp]
\center{
\includegraphics[width=4in]{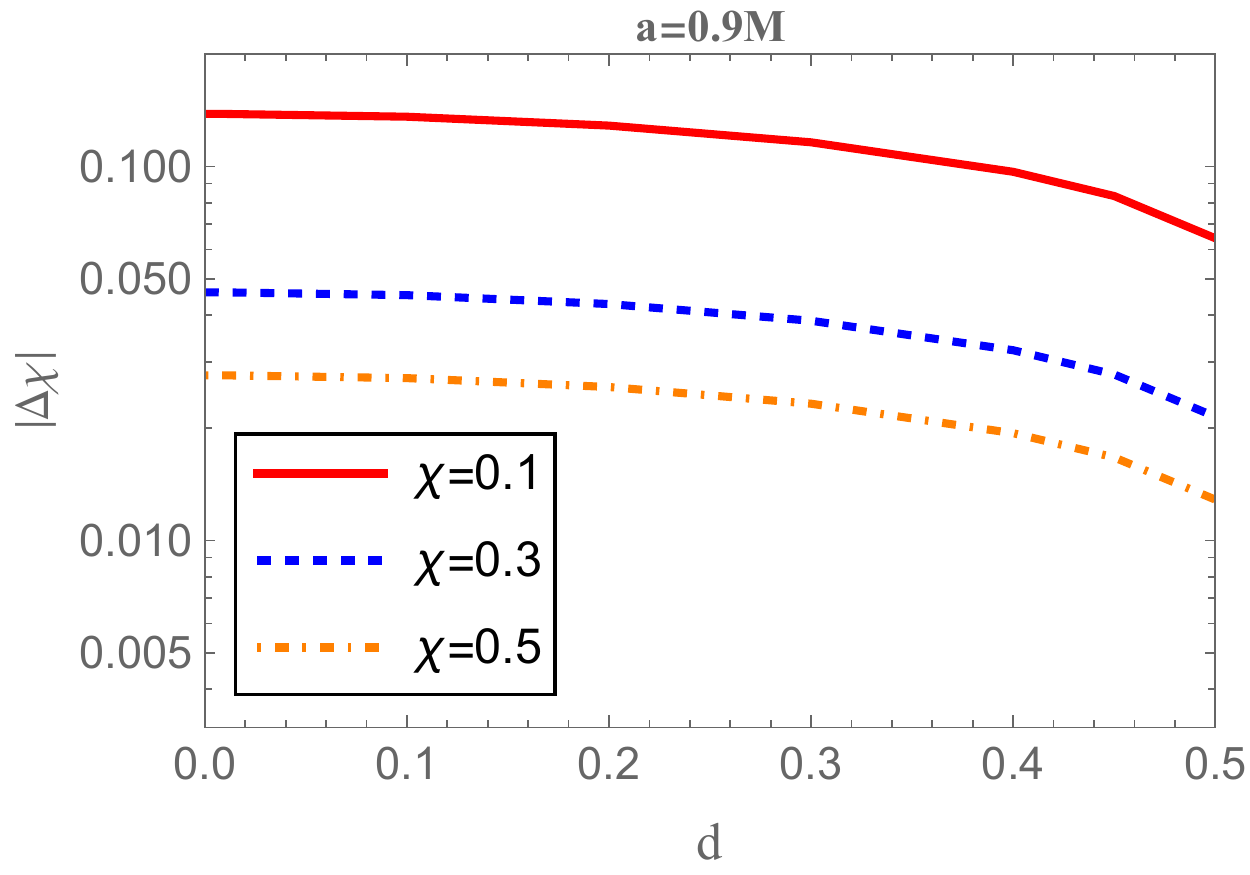}
\caption{Fixing $a=0.9M$, we show the spin resolution as a function of the scalar charge $d$ for different secondary spin $\chi=0.1, 0.3, 0.5$ as an example.}\label{fg:resolution}}
\end{figure}

Furthermore, the presence of scalar field amplifies the dephasing, which leads to systematic improvement in the spin resolution of this model \cite{Piovano:2020zin}. 
Considering two waveforms that differ only by the secondary spin $\chi_1$ and $\chi_2$. 
We can evaluate the minimum detectable spin difference by the phase resolution \cite{Piovano:2020ooe,Piovano:2020zin}
\begin{equation}\label{eq:spinresolution}
|\Delta \chi|=|\chi_1-\chi_2|>\frac{\Delta\varphi}{\left|\delta \varphi_{\mathrm{GW}}\right|},
\end{equation}
where $\Delta\varphi=1$ rad has been constrained eailer, and $\delta\varphi_{\mathrm{GW}}$ can be replaced by $N_{\chi}^d-N_{\chi=0}^d$ in this model. 
Fig. \ref{fg:resolution} shows that the spin resolution is effectively improved by the scalar charge. 
As the scalar charge increases to $d=0.5$, the improvement exceeds $100\%$. 
Moreover, we observe that the slope of the resolution becomes more skewed as $d$ increases to beyond $d\approx 0.45$. 
This suggests that a larger scalar charge is more beneficial for improving the spin resolution.

%%%%%%%%%%%%%%%%%%%%
%%%%%%%%%%%%%%%%%%%%

Previously, we derived exact results on how the presence of scalar field amplifies the detection of the secondary spin in terms of detection limit and spin resolution. 
We now turn our attention to assessing the detection capabilities of space-based GW detectors, particularly LISA, Taiji and TianQin. 
Additional details on the calculations and related detector configurations are provided in Appendix \ref{sec=params}. 
In this setup, the secondary body has a mass of $m_p=10M_{\odot}$ with a scalar charge of $d=0.5$, while the primary mass is $M=4\times10^5M_{\odot}$ with spin $a=0.9M$. 

One way to quantitatively assess the detectability of a GW detector is through the faithfulness $\mathcal{F}$, which compares two GW signals with and without the presence of the secondary spin. 
The faithfulness measures the difference between these two signals weighted by the noise spectral density of the GW detector. 
For example, with a signal-noise ratio(SNR) $\rho=30$, the capability of GW detector requires faithfulness $\mathcal{F}\leq 0.988$ to determine the parameter resolution of the secondary spin. 
Here we calculate the faithfulness as a function of the secondary spin for two scalar charge values, $d=0.001$ and $d=0.5$, respectively. 

%%%%%%%%%%%%%%%%%%%%
\begin{figure}[thbp]
\center{
\includegraphics[width=2.1in]{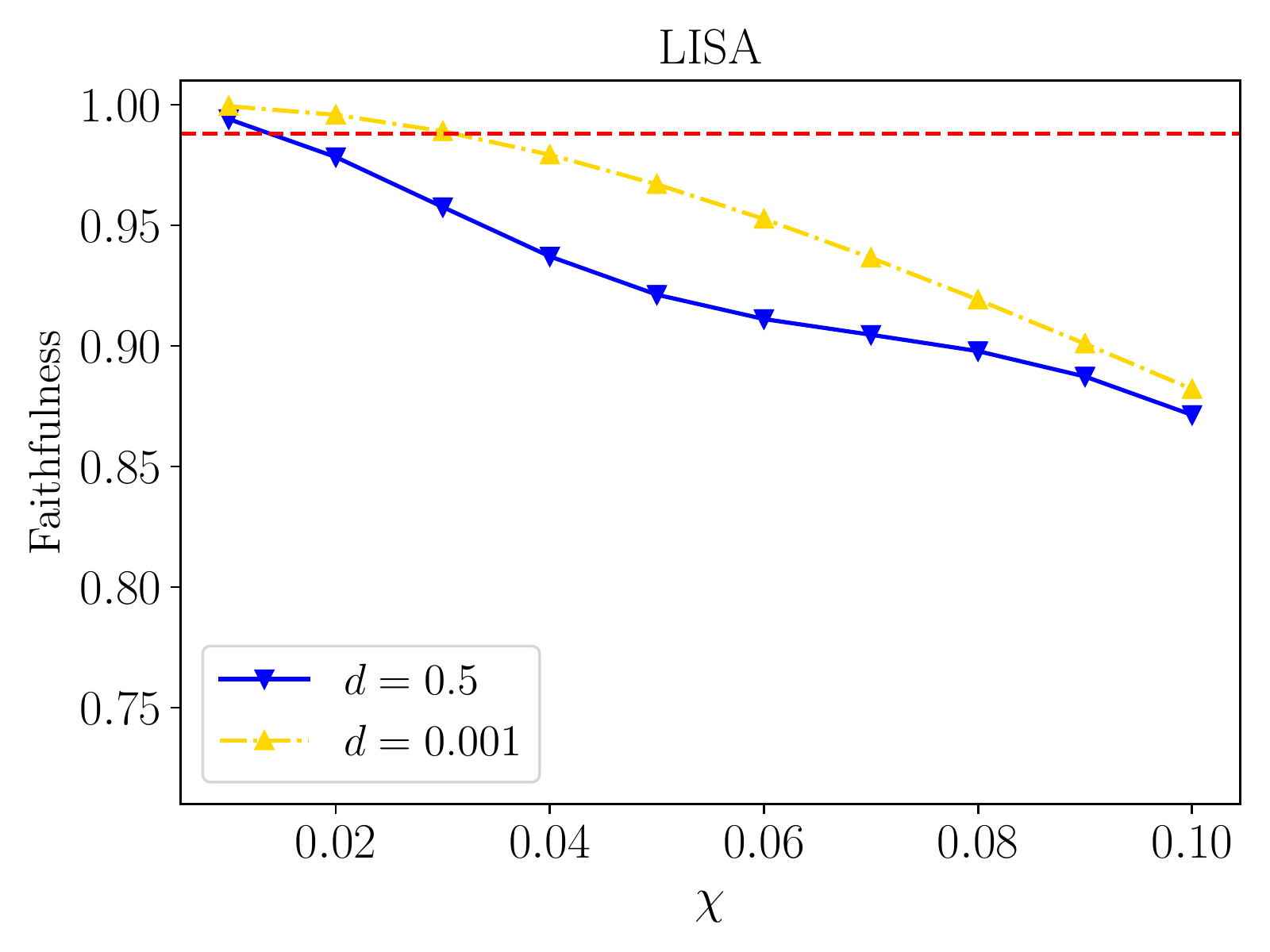}
\includegraphics[width=2.1in]{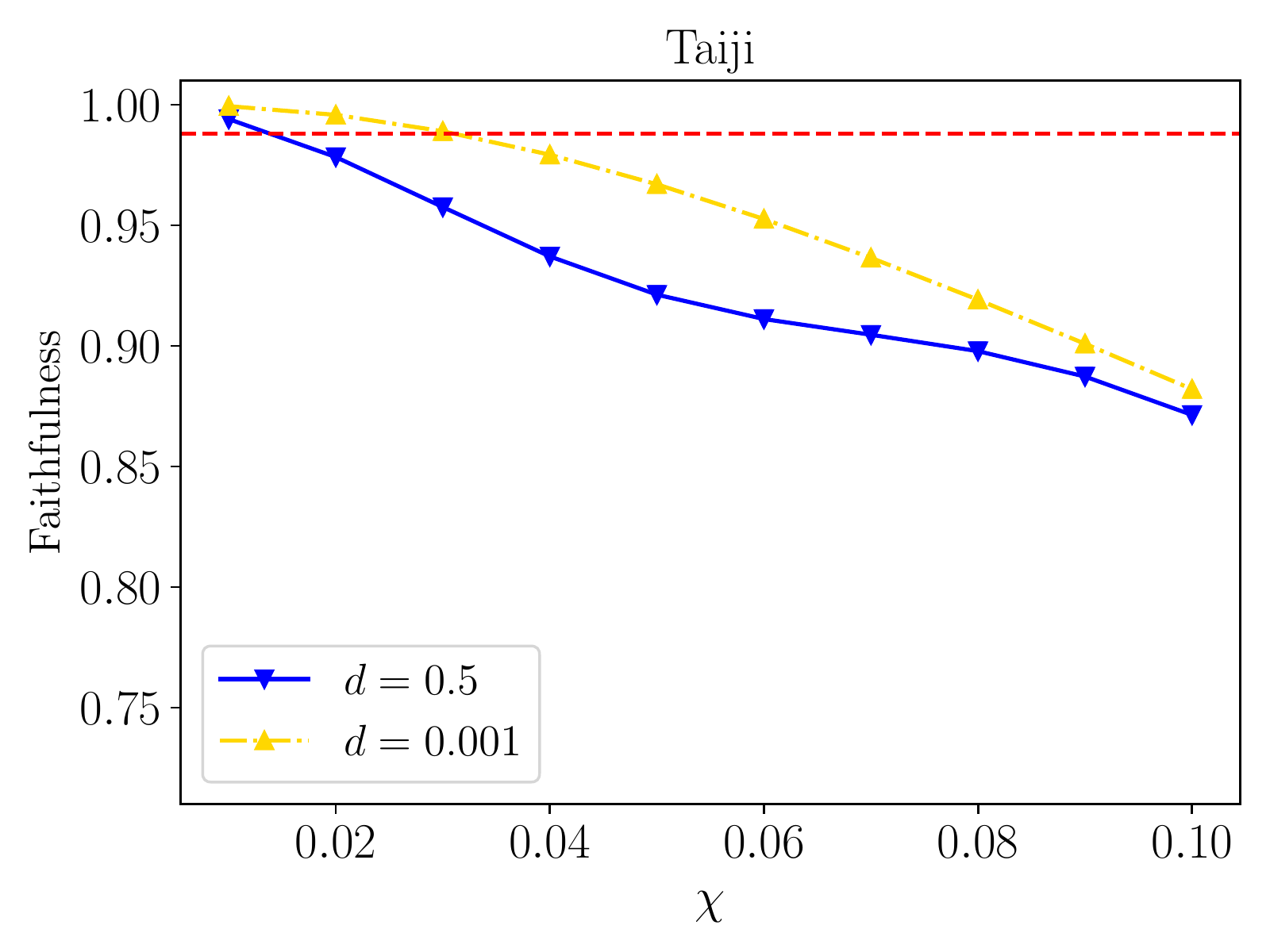}
\includegraphics[width=2.1in]{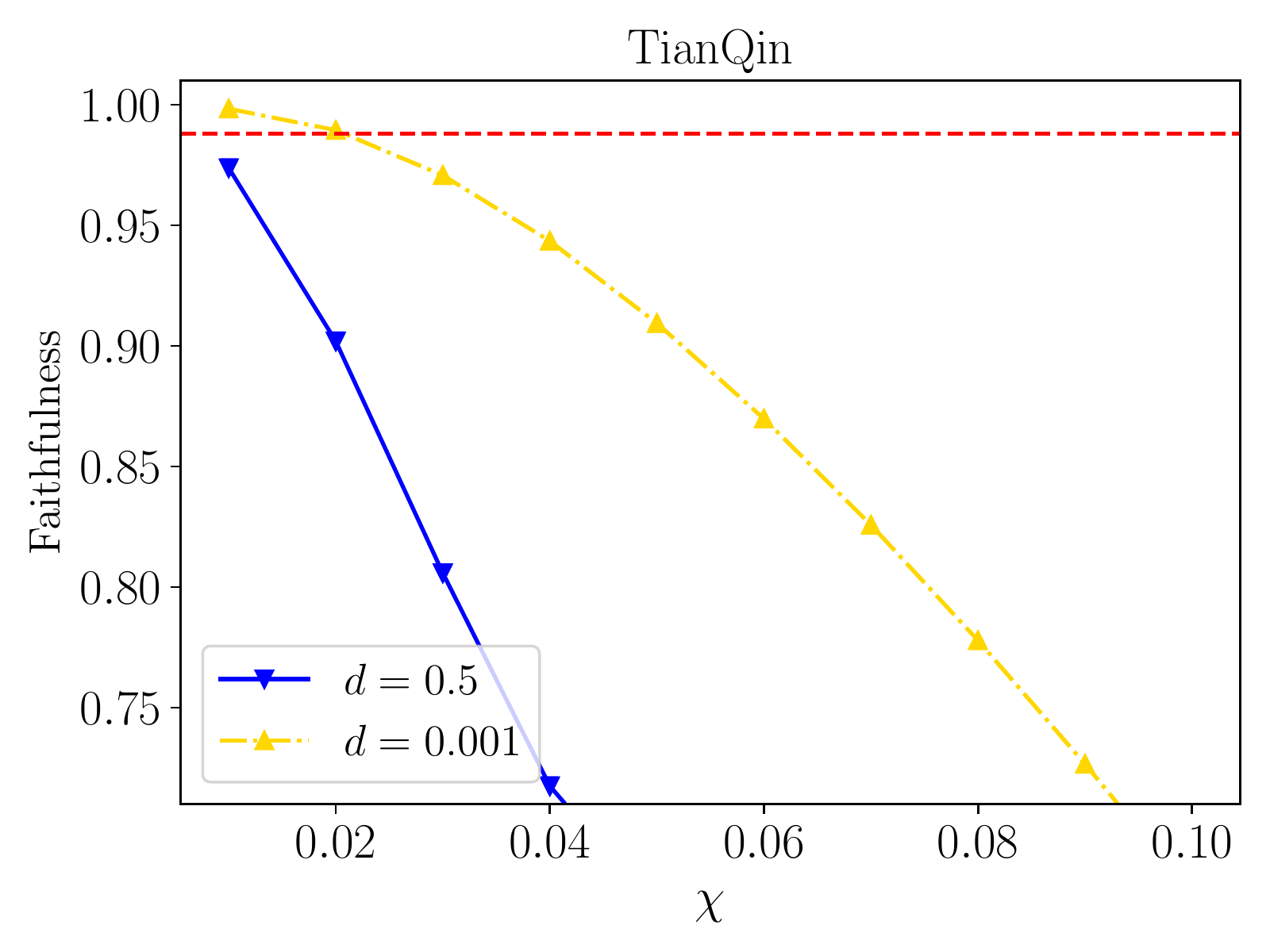}
\caption{The faithfulness as the function of the secondary spin with $d=0.5$ and $d=0.001$ for LISA, Taiji and TianQin, respectively. Here the parameters are set as $a=0.9M$, $M=4\times10^5M_\odot$, $m_p=10M_\odot$ and $r_{start}=11.53M$ with one-year evolution.}\label{fg:faithfulness}}
\end{figure}
%%%%%%%%%%%%%%%%%%%%

The results, illustrated in Fig. \ref{fg:faithfulness}, demonstrate that after one-year evolution, the faithfulness decreases with increasing secondary spin for all three GW detectors LISA, Taiji and TianQin. 
In most regions of the secondary spin, the faithfulness is sufficiently small for all three GW detectors to distinguish between the GW signals with and without a secondary spin. 
Interestingly, the value of faithfulness for $d=0.5$ is consistently lower than those for $d=0.001$, indicating that the GW signal for $d=0.5$ is generally more favorable than that for $d=0.001$. 
By setting the threshold at $\mathcal{F}=0.988$, we can observe that the existence of the scalar charge $d$ improves the resolution of the secondary spin. 
For example, for the GW detector TianQin, the resolution improves from $\chi=0.025$ when $d=0.001$ to $\chi<0.01$ when $d=0.5$. 
These results support our conclusion that the presence of the scalar field enhances the detectability of the secondary spin and improves the resolution of the secondary spin for space-based detectors.

\begin{figure}[thbp]
\center{
\includegraphics[width=2.1in]{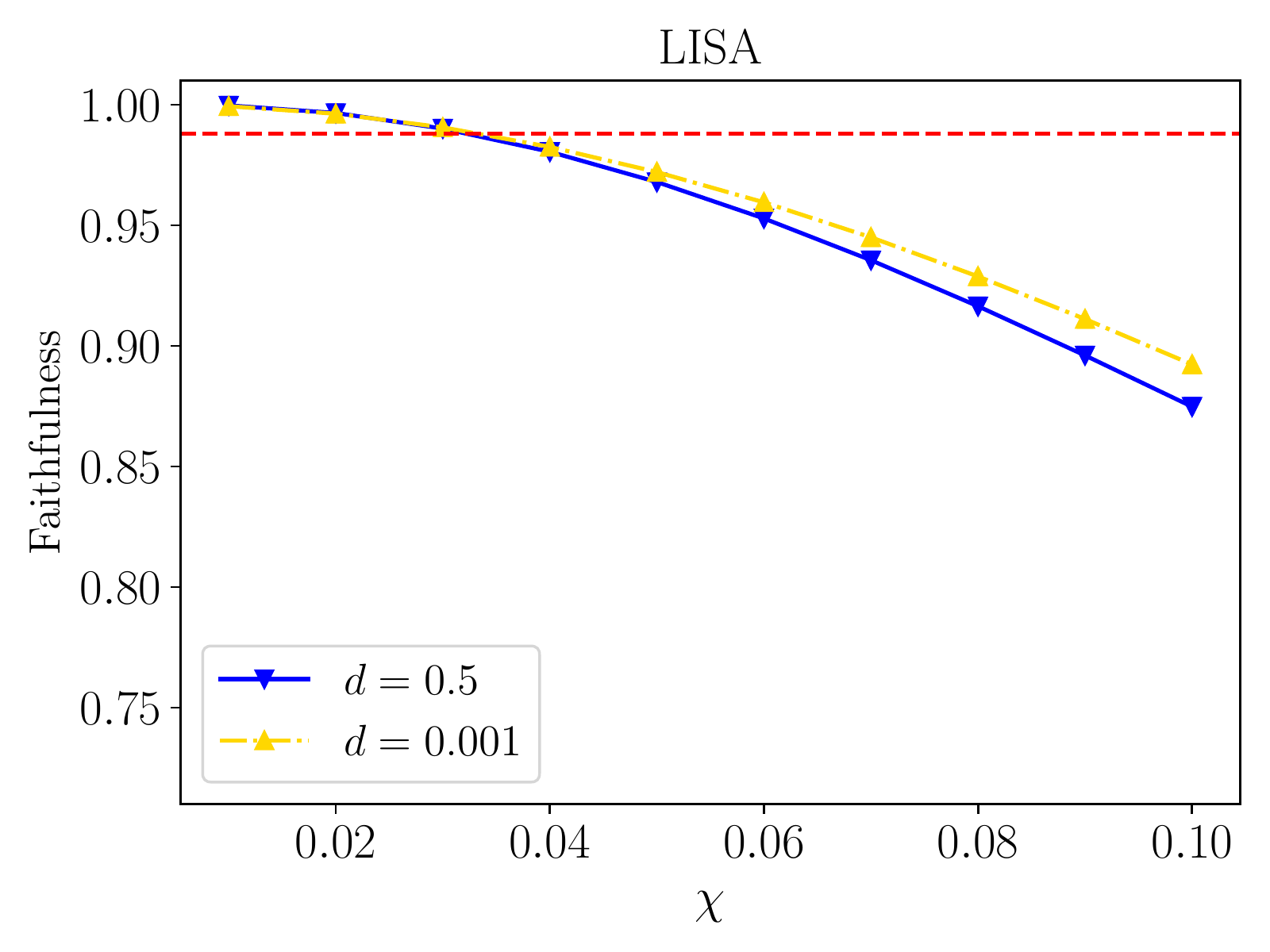}
\includegraphics[width=2.1in]{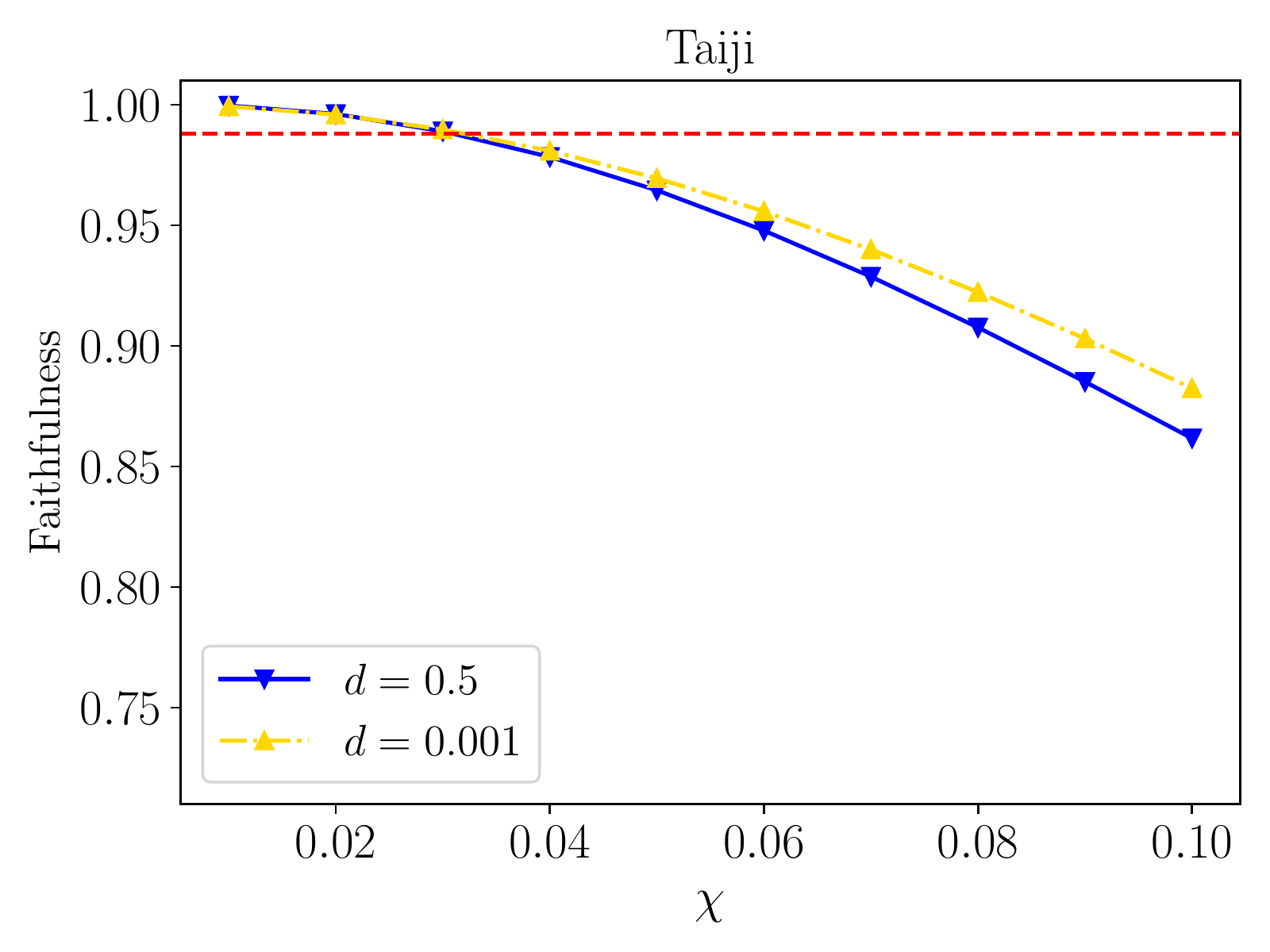}
\includegraphics[width=2.1in]{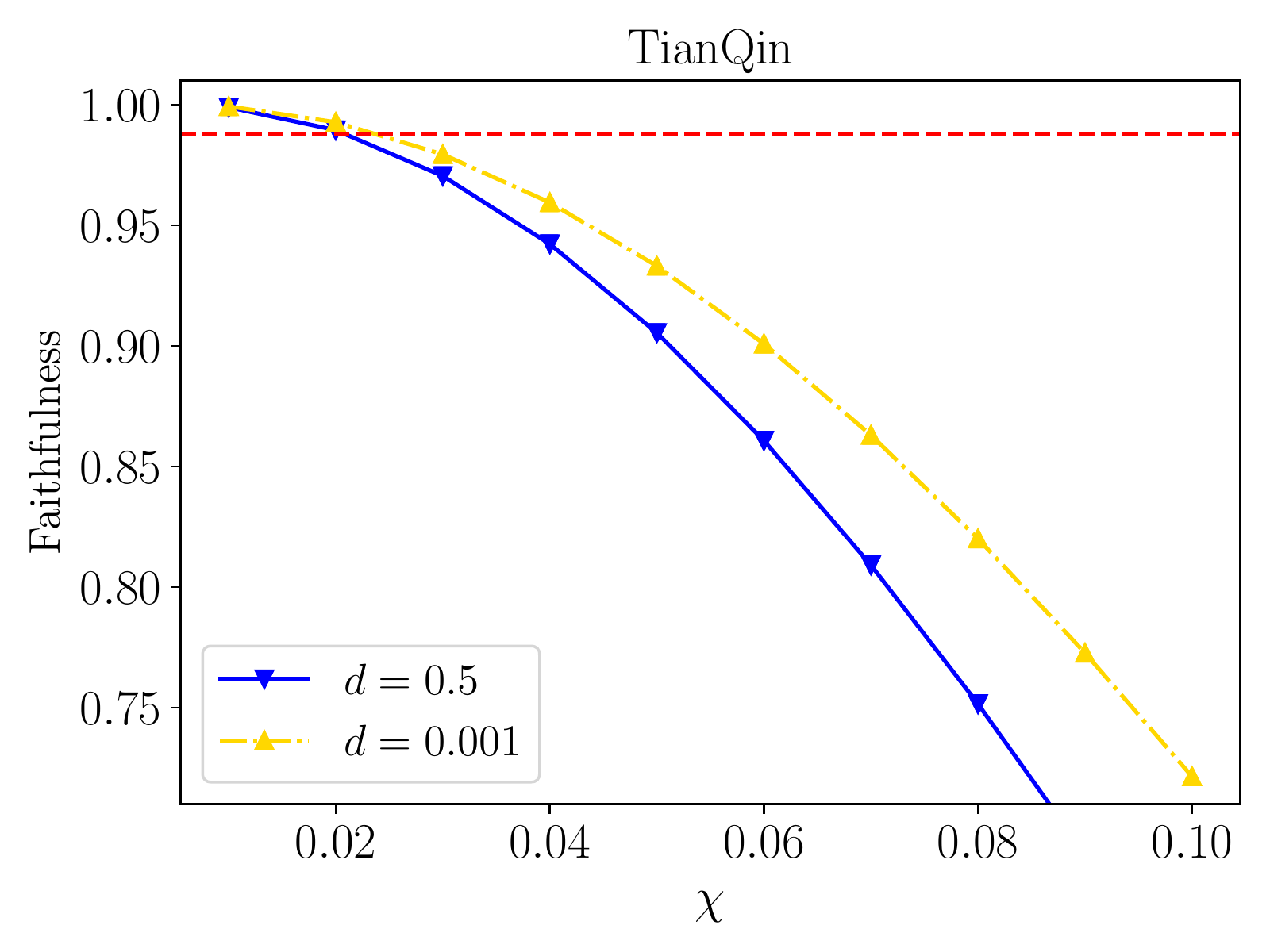}
\caption{The faithfulness as the function of the secondary spin with $d=0.5$ and $d=0.001$ for LISA, Taiji and TianQin, respectively. Here the parameters are set as $a=0.9M$, $M=1\times10^6M_\odot$, $m_p=10M_\odot$ and $r_{start}=7.2M$ with one-year evolution.}\label{fg:faithfulness2}}
\end{figure}

\begin{figure}[thbp]
\center{
\includegraphics[width=2.1in]{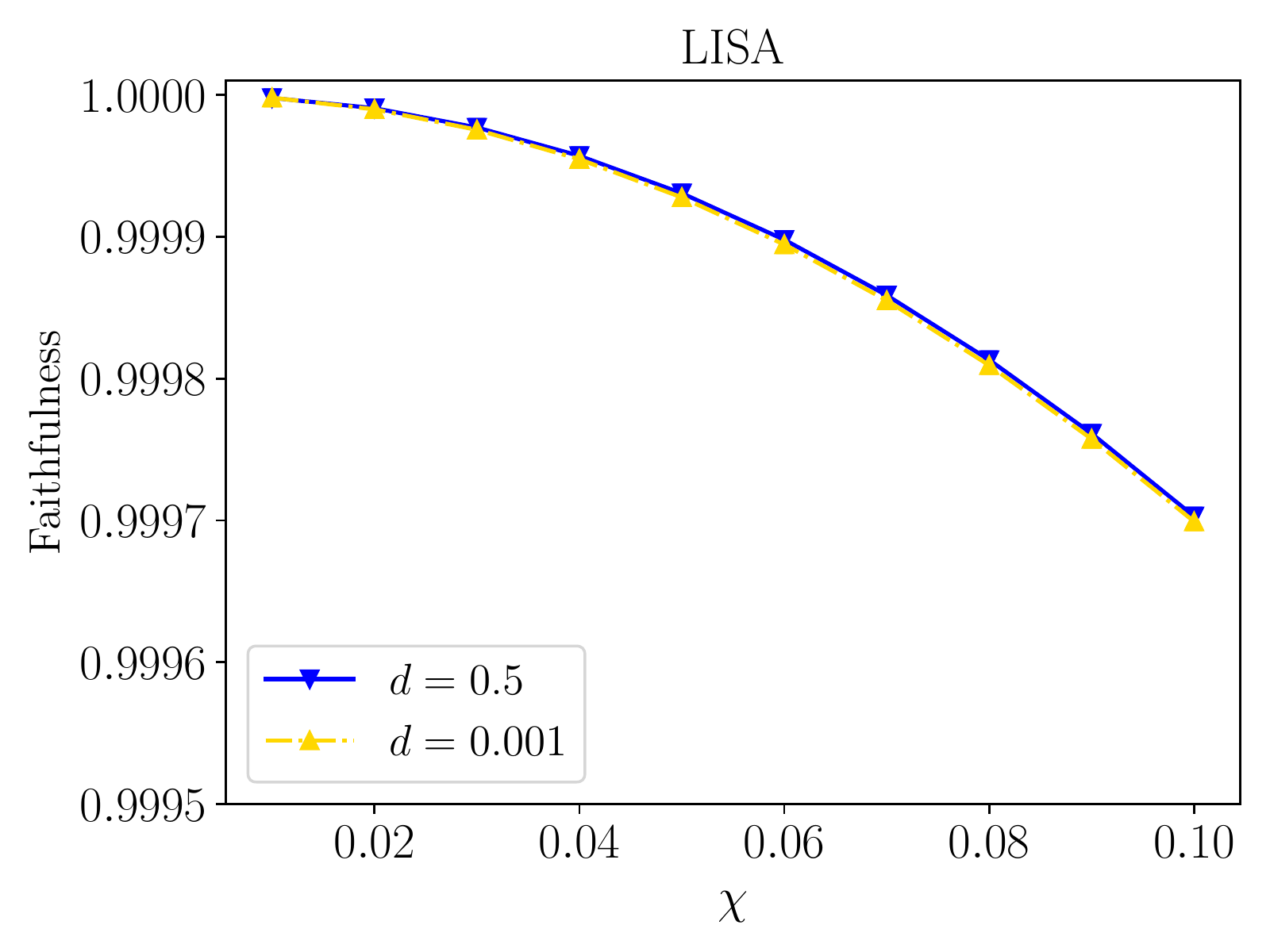}
\includegraphics[width=2.1in]{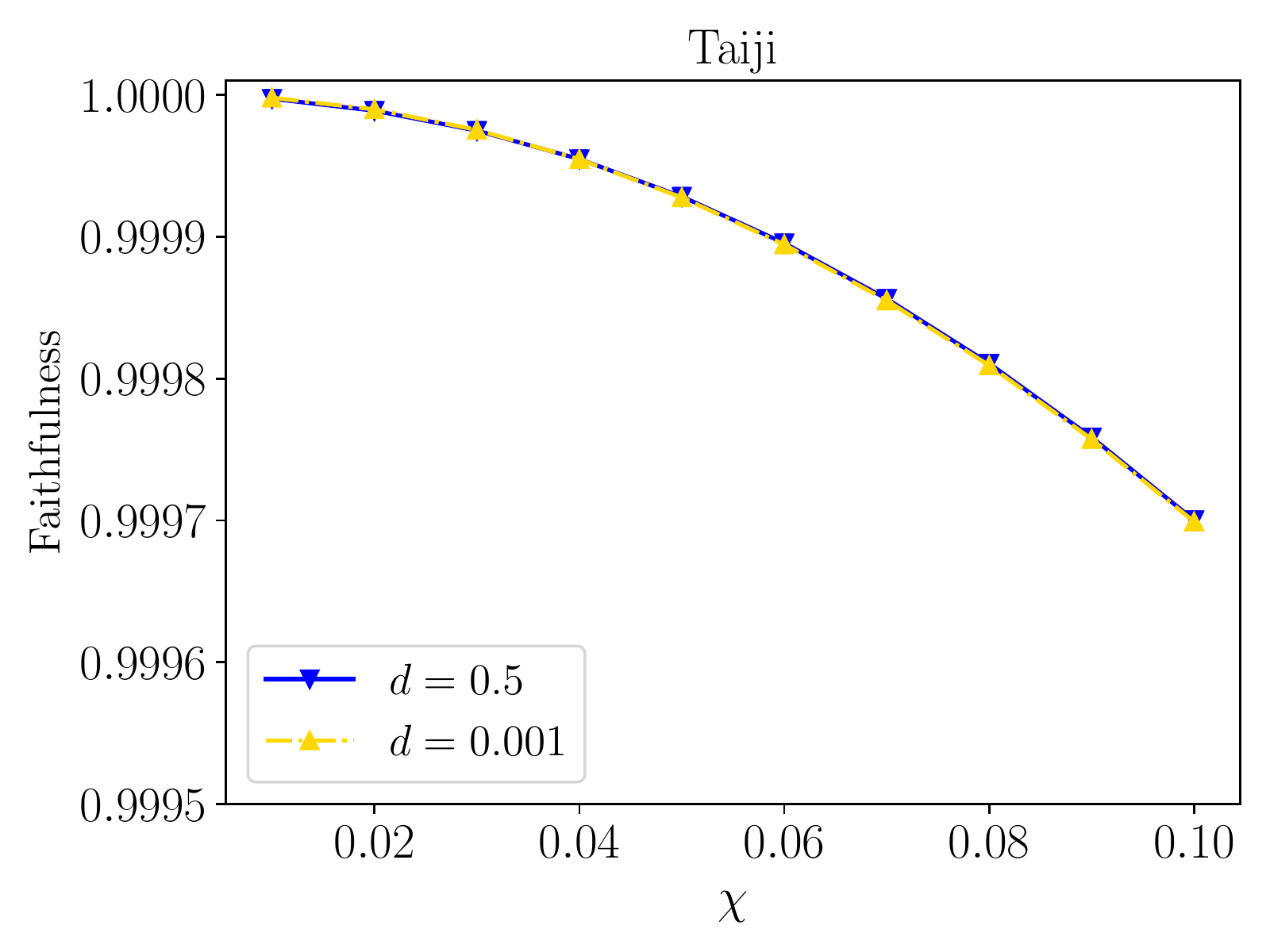}
\includegraphics[width=2.1in]{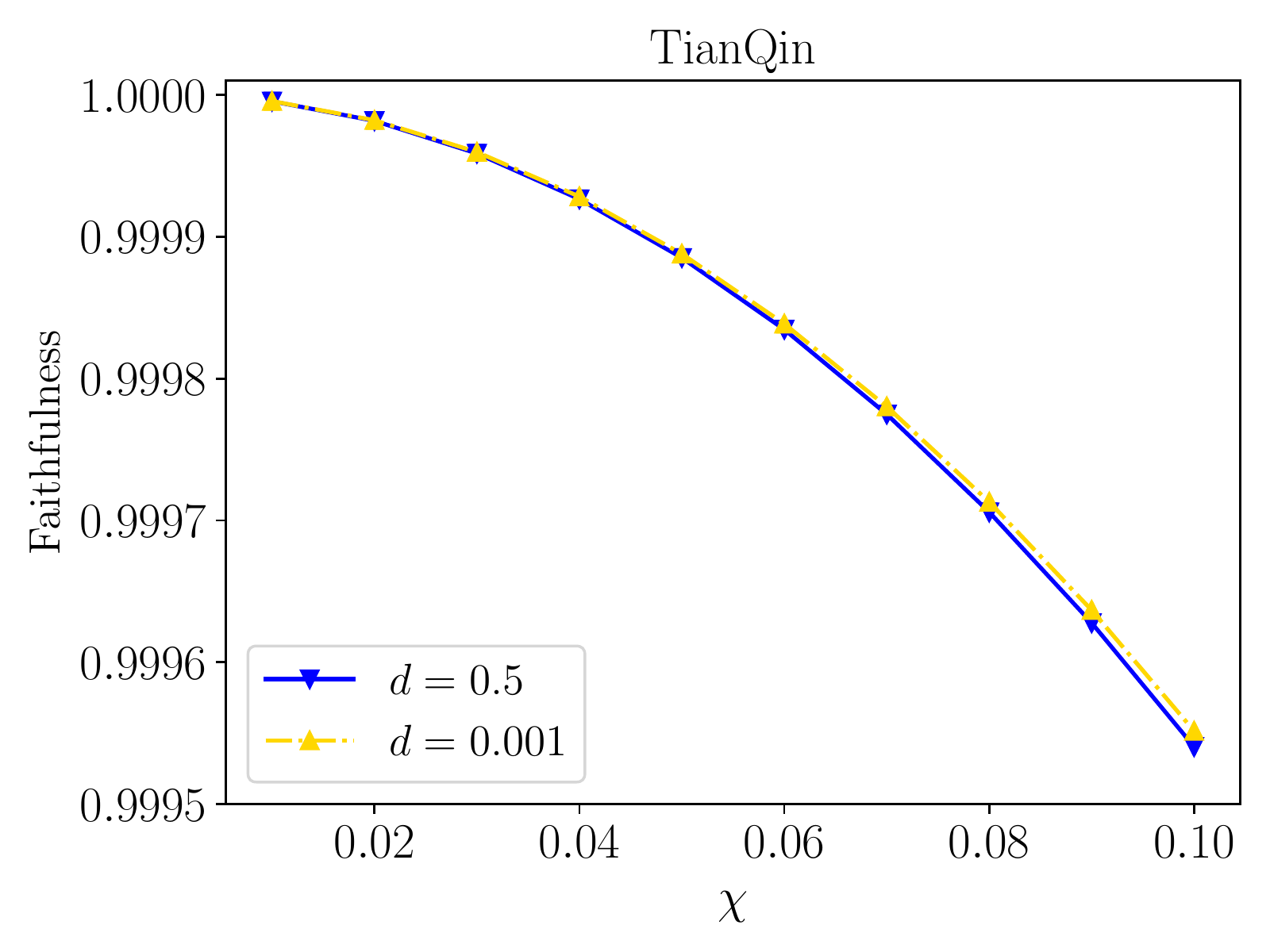}
\caption{The faithfulness as the function of the secondary spin with $d=0.5$ and $d=0.001$ for LISA, Taiji and TianQin, respectively. Here the parameters are set as $a=0.9M$, $M=1\times10^7M_\odot$, $m_p=10M_\odot$ and $r_{start}=3.0M$ with one-year evolution.}\label{fg:faithfulness3}}
\end{figure}

Moreover, when considering $M=4\times 10^{5}M_{\odot}$, the faithfulness of TianQin is found to be better than the other two GW detectors LISA and Taiji, as shown in Fig.\ref{fg:faithfulness}. 
This is expected, since the TianQin detector exhibits higher sensitivity in the high-frequency range \cite{Gong:2021gvw}. 
In addition, we observe an increase in faithfulness with the growth of the primary BH mass when  comparing the faithfulness of the three space-based GW detectors with primary mass of $M=1\times10^6M_\odot$ and $M=1\times10^7M_\odot$ in Fig. \ref{fg:faithfulness2} and Fig. \ref{fg:faithfulness3}, respectively.
Notably, for $M\times10^6M_\odot$, the presence of the scalar charge $d$ has little effect on improving the resolution of the secondary spin $\chi$, whereas for $M=1\times10^7M_\odot$, the secondary spin $\chi$ is indistinguishable from GR.
This implies that scalar radiation is comparatively more efficient in the far-field zone than in the near-field zone, as illustrated in Fig.\ref{fg:fluxx}, when compared to gravitational radiation.
%This implies that scalar radiation is comparatively more efficient in the far-field zone than in the near-field zone, as compared to gravitational radiation. as demonstrated in Fig. \ref{fg:fluxx}. 
When the mass of the primary BH is moderate, the evolution of the secondary begins far away from the primary BH. 
However, when the mass of the primary BH is considerable, the one-year evolution of the secondary occurs in the near-field zone, as noted in the captions of Fig. \ref{fg:faithfulness2} and Fig. \ref{fg:faithfulness3}.
In conclusion, the secondary spin $\chi$ is more suitable for detection in the region when the mass of the primary BH $M$ is not large, and TianQin is the optimal choice for detection of the secondary spin.

%%%%%%%%%%%%%%%%%%%%
%%%%%%%%%%%%%%%%%%%%

%%%%%%%%%%%%%%%%%%%%
%%%%%%%%%%%%%%%%%%%%

%%%%%%%%%%%%%%%%%%%%%%%%%%%%%%%%%%%%%%%%%%%%%%%%%%%%%%%%%%%%%%%%%%%%%%%%%%%%
\section{Further discussions and concluding remarks}\label{sec=conclusion}

In this paper, we discuss the detectability of the secondary spin in the EMRI system within a modified gravity model coupled with a scalar field. 
The central BH, which reduces to a Kerr one, is circularly spiraled by a scalar-charged spinning secondary body on the equatorial plane.  
In contrast to GR, the presence of scalar field supports an additional radiation channel for GW, offering a modified GW template that could potentially shed light on the properties of binary systems.

This model considers a one-year adiabatic evolution starting at $r_{start}=11.53M$ for the Kerr BH with primary spin $a=0.9M$ and mass $M=4\times10^5M_{\odot}$, while the secondary has a mass of $m_p=10M_{\odot}$. 
By numerically solving the inhomogeneous Teukolsky equation and scalar perturbation equation, we calculate the total energy fluxes and dephasing for a range of the model parameters including $d\in[0,0.5]$ and $\chi\in[0,0.5]$.
Our analysis of the orbital evolution and total energy fluxes confirms that the secondary spin plays a relatively secondary role in the EMRI system, suggesting a limited influence on detecting the scalar charge.
Nonetheless, we have determined that the inclusion of scalar radiation may enhance the detection threshold and spin resolution, as evidenced by our analysis of its overall dephasing.
%However, we find that the additional scalar radiation could help improve the detection limit and  spin resolution by studying the total dephasing of this model.
As shown in Fig. \ref{fg:dephase}, the increasing of the scalar charge leads to a lower detection limit, with an improvement from $\chi=0.014$ for $d=0$ to $\chi=0.006$ for $d=0.5$.
Moreover, the spin resolution, determined by Eq. \eqref{eq:spinresolution}, is enhanced over $100\%$ if the scalar charge is increased to $d=0.5$. 
A more pronounced tilt in the spin resolution slope, as depicted in Fig. \ref{fg:resolution}, corroborates  that a greater scalar charge confers a more substantial enhancement to the spin resolution.
%The more skewed slope of the spin resolution in Fig. \ref{fg:resolution} supports the conclusion that a larger scalar charge is more beneficial for improving the spin resolution.

To validate our theoretical analysis, we calculate the faithfulness to compare GW signals with spinless results.
It is proved that the presence of the scalar field enhances the detectability of the secondary spin by all three space-based GW detectors.
In Fig. \ref{fg:faithfulness}, our results show that the faithfulness decreases with the growth of the scalar charge and is sufficiently small to distinguish the GW signals from GR in most regions of the secondary spin.
Moreover, the value of faithfulness for $d=0.5$ is always lower than that for $d=0.001$, indicating an improved spin resolution that will be more precise in regions with large secondary spin.
The secondary spin detection limit of space-based detectors can be determined by the threshold at the faithfulness $\mathcal{F}=0.988$, which is found to be improved by the scalar charge. For TianQin, as an example, the detection limit improves from $\chi=0.025$ when $d=0.001$ to $\chi<0.01$ when $d=0.5$.

Furthermore, our results also show that TianQin is a better choice for detecting the secondary spin in this model, as its behavior of faithfulness is better than LISA and Taiji when considering the primary mass $M=4\times10^5M_\odot$.
This is because TianQin has greater sensitivity in the high-frequency region. 
However, as we increase the primary mass, the presence of the scalar charge has little effect on improving the resolution of the secondary spin when $M=1\times10^6M_\odot$ as shown in Fig. \ref{fg:faithfulness2}.
Finally, the secondary spin is indistinguishable from GR when $M=1\times10^7M_\odot$ as shown in Fig. \ref{fg:faithfulness3}.
This is reasonable because the scalar radiation is more effective than gravitational radiation in the far-field zone, but the whole one-year evolution is completed in the near-field zone where gravitational radiation grows faster \cite{Maselli:2020zgv, Guo:2022euk}.

In summary, our study investigated the detectability of the secondary spin in the modified gravity coupled with a scalar field. 
We found that the presence of the scalar field amplifies the secondary spin effect, allowing for the detection of a lower limit value of secondary spin and an improved resolution of secondary spin detection when the scalar charge is sufficiently large.
Our findings suggest that the secondary spin is more suitable for detection when the primary mass is not large, and TianQin is the optimal choice for detection.

The implications of our results are crucial for future observations of EMRIs and for testing modified gravity theories in the strong field regime.
It suggests that the presence of scalar fields could substantially impact the dynamics of compact objects in the vicinity of supermassive black holes, leading to important consequences for interpreting GW signals.
Moreover, our study supports that the EMRI model in modified gravity theories can enable investigations onto the properties of binaries.
This is an important direction for future research, as alternative theories of gravity may offer a better tool for exploring the universe than GR. 
Consequently, we can further discuss constraints on cosmological parameters \cite{MacLeod:2007jd, Laghi:2021pqk} and detection of model parameters \cite{Huerta:2008gb,Gair:2012nm}. 
In addition, the interaction between the secondary spin and the scalar field should be considered for further precise discussion on the detection of the secondary spin, which we assumed away the simplest case.
Lastly, other modified gravities may also have additional radiations, making studying EMRI systems in these models valuable for future research.

%\newpage

\begin{acknowledgments}
This research is supported by the National Key Research and Development Program of China under Grant No. 2020YFC2201400. 
YG acknowledges the support by the National Natural Science Foundation of China under Grant Nos. 11875136 and 12147120.
\end{acknowledgments}

\appendix
%%%%%%%%%%%%%%%%%%%%%%%%%%%%%%%%%%%%%%%%%%%%%%%%%%%%%%%%%%%%%%%%%%%%%%%%%%%%

\section{The dephasing data}\label{sec=data}

In this section, we show the dephasing data of our model.
The model parameters are set by $m_p=10M_\odot$, $M=4\times10^5M_\odot$, $a=0.9M$.
Table \ref{tab:method1} show the dephasing $|N_{\chi}^d-N_{\chi=0}^d|$ by the modified data processing approach by setting $r_{start}=11.53M$ with one-year evolution.
Table \ref{tab:method2} show the dephasing $|N_{\chi}^d-N_{\chi=0}^d|$ by the original approach with one-year evolution before the plunge into the ISCO. 
Table \ref{tab:absphasing} show the residual dephasing between our total GW phase and the summation of all contributions from scalar charge and secondary spin.

%%%%%%%%%%%%%%%%%%%%
\begin{table}[h]
\centering
\caption{The table of dephasing $|N_{\chi}^d-N_{\chi=0}^d|$ with different secondary spin $\chi$ and scalar charge $d$. Here we use the modified data processing method by setting $r_{start}=11.53M$ with one-year evolution.}
\label{tab:method1}
\begin{tabular}{|l|l|l|l|l|l|l|l|l|l|}
\hline
dephasing &
  \multicolumn{1}{c|}{d=0} &
  d=0.001 &
  d=0.01 &
  \multicolumn{1}{c|}{d=0.1} &
  \multicolumn{1}{c|}{d=0.2} &
  \multicolumn{1}{c|}{d=0.3} &
  \multicolumn{1}{c|}{d=0.4} &
  d=0.45 &
  \multicolumn{1}{c|}{d=0.5} \\ \hline
$\chi$=0     & 0      & 0      & 0      & 0      & 0      & 0      & 0      & 0      & 0      \\ \hline
$\chi$=0.01  & 0.72382  & 0.72382  & 0.72394  & 0.73645  & 0.77790  & 0.86221  & 1.03381  & 1.20007    & 1.55652  \\ \hline
$\chi$=0.016 & 1.15811  & 1.15811  & 1.15831  & 1.17833  & 1.24463  & 1.37952  & 1.65410  & 1.92011   & 2.49041   \\ \hline
$\chi$=0.018 & 1.30288  & 1.30288  & 1.30310  & 1.32562  & 1.40021    & 1.55196  & 1.86086  & 2.16012   & 2.80171  \\ \hline
$\chi$=0.02  & 1.44764  & 1.44764  & 1.44789  & 1.47291  & 1.55579  & 1.72440  & 2.06762  & 2.40013    & 3.113  \\ \hline
$\chi$=0.06  & 4.34288  & 3.34289  & 3.34362  & 4.41867  & 4.66732  & 5.17315  & 6.20275  & 7.20023    & 9.33856  \\ \hline
$\chi$=0.1   & 7.23806  & 7.23808  & 7.23930  & 7.36438  & 7.77878  & 8.62181  & 10.3377 & 12.0001   & 15.5636 \\ \hline
$\chi$=0.2   & 14.4748 & 14.4758 & 14.4783 & 14.7284 & 15.5572 & 17.2431 & 20.6746 & 23.9988 & 31.1235 \\ \hline
$\chi$=0.3   & 21.7132 & 21.7132 & 21.7169 & 22.0921 & 23.3351 & 25.8638 & 31.0105 & 35.9961 & 46.6799  \\ \hline
$\chi$=0.4   & 28.9502  & 28.9503  & 28.9552 & 29.4554 & 31.1127 & 34.4840 & 41.3456 & 47.992 & 62.2327 \\ \hline
$\chi$=0.5   & 36.1869 & 36.187 & 36.1931 & 36.8184 & 38.8898  & 43.1037 & 51.6798  & 59.9865 & 77.8919 \\ \hline
\end{tabular}
\end{table}
%%%%%%%%%%%%%%%%%%%%

%%%%%%%%%%%%%%%%%%%%
\begin{table}[h]
\centering
\caption{The table of dephasing $|N_{\chi}^d-N_{\chi=0}^d|$ with different secondary spin $\chi$ and scalar charge $d$. Here we use the data processing method with one-year evolution before the plunge into the ISCO.}
\label{tab:method2}
\begin{tabular}{|l|l|l|l|l|l|l|l|l|l|}
\hline
dephasing &
  \multicolumn{1}{c|}{d=0} &
  d=0.001 &
  d=0.01 &
  \multicolumn{1}{c|}{d=0.1} &
  \multicolumn{1}{c|}{d=0.2} &
  \multicolumn{1}{c|}{d=0.3} &
  \multicolumn{1}{c|}{d=0.4} &
  d=0.45 &
  \multicolumn{1}{c|}{d=0.5} \\ \hline
$\chi$=0     & 0        & 0        & 0        & 0        & 0        & 0        & 0        & 0        & 0        \\ \hline
$\chi$=0.01  & 0.25049 & 0.25049 & 0.25049 & 0.25048 & 0.25044 & 0.25038 & 0.25030 & 0.25025 & 0.25019  \\ \hline
$\chi$=0.016 & 0.40078 & 0.40078 & 0.40078 & 0.40076 & 0.40071 & 0.40061 & 0.40048 & 0.40040 & 0.40030 \\ \hline
$\chi$=0.018 & 0.45088 & 0.45088 & 0.45088 & 0.45086 & 0.45080 & 0.45069 & 0.45054 & 0.45045 & 0.45034 \\ \hline
$\chi$=0.02  & 0.50098 & 0.50098 & 0.50098 & 0.50095 & 0.50088 & 0.50077 & 0.50060 & 0.50050 & 0.50038  \\ \hline
$\chi$=0.06  & 1.50293  & 1.50293  & 1.50293  & 1.50286  & 1.50265  & 1.5023   & 1.50179  & 1.50149  & 1.50114  \\ \hline
$\chi$=0.1   & 2.50488  & 2.50488  & 2.50488  & 2.50477  & 2.50442  & 2.50383  & 2.50299  & 2.50248  & 2.5019   \\ \hline
$\chi$=0.2   & 5.00977  & 5.00977  & 5.00977  & 5.00954  & 5.00884  & 5.00766  & 5.00599  & 5.00496  & 5.0038   \\ \hline
$\chi$=0.3   & 7.51466  & 7.51466  & 7.51466  & 7.51431  & 7.51326  & 7.51149  & 7.50898  & 7.50744  & 7.50571  \\ \hline
$\chi$=0.4   & 10.0196  & 10.0196  & 10.0196  & 10.0191  & 10.0177  & 10.0153  & 10.012   & 10.0099  & 10.0076  \\ \hline
$\chi$=0.5   & 12.5245  & 12.5245  & 12.5245  & 12.5239  & 12.5221  & 12.5192  & 12.515   & 12.5124  & 12.5095  \\ \hline
\end{tabular}
\end{table}
%%%%%%%%%%%%%%%%%%%%

%%%%%%%%%%%%%%%%%%%%
\begin{table}[h]
\centering
\caption{The table of dephasing $|N_{\chi}^d-N_{\chi=0}^d-N^{d=0}_{\chi}+N^{d=0}_{\chi=0}|$ with different secondary spin $\chi$ and scalar charge $d$. Here we use the modified data processing method by setting $r_{start}=11.53M$ with one-year evolution.}
\label{tab:absphasing}
\begin{tabular}{|l|l|l|l|l|l|l|l|l|l|}
\hline
dephasing &
  \multicolumn{1}{c|}{d=0} &
  \multicolumn{1}{c|}{d=0.001} &
  \multicolumn{1}{c|}{d=0.01} &
  \multicolumn{1}{c|}{d=0.1} &
  \multicolumn{1}{c|}{d=0.2} &
  \multicolumn{1}{c|}{d=0.3} &
  \multicolumn{1}{c|}{d=0.4} &
  \multicolumn{1}{c|}{d=0.45} &
  \multicolumn{1}{c|}{d=0.5} \\ \hline
$\chi$=0     & 0      & 0      & 0      & 0      & 0      & 0      & 0      & 0      & 0      \\ \hline
$\chi$=0.01  & 0  & $4.30\times10^{-7}$  & $4.35\times10^{-5}$  & 0.00439  & 0.018124 & 0.043099 & 0.08337  & 1.11160& 0.14744    \\ \hline
$\chi$=0.016 & 0  & $6.93\times10^{-7}$  & $6.95\times10^{-5}$  & 0.00703  & 0.02300  & 0.06896  & 0.13340  & 0.17856  & 0.23591  \\ \hline
$\chi$=0.018 &0.  & $7.78\times10^{-7}$   & $7.82\times10^{-5}$  & 0.00790  & 0.03262  & 0.07758  & 0.15007  & 0.20088  & 0.26540  \\ \hline
$\chi$=0.02  & 0  & $8.66\times10^{-7}$  & $8.69\times10^{-5}$  & 0.00878  & 0.03625  & 0.08620  & 0.16674  & 0.22320  & 0.29489  \\ \hline
$\chi$=0.06  & 0  & $2.61\times10^{-6}$  & $2.61\times10^{-4}$  & 0.02634  & 0.10874  & 0.25859  & 0.50023  & 0.66959  & 0.88465  \\ \hline
$\chi$=0.1   & 0  & $4.35\times10^{-6}$  & $4.35\times10^{-4}$  & 0.04391  & 0.18124  & 0.43097  & 0.83370  & 1.11597  & 1.47439  \\ \hline
$\chi$=0.2   & 0  & $8.69\times10^{-6}$   & $8.69\times10^{-4}$  & 0.08781  & 0.36246  & 0.86192  & 1.66734  & 2.23187  & 2.94867  \\ \hline
$\chi$=0.3   & 0  & $1.30\times10^{-5}$  & 0.001304  & 0.13171  & 0.54368  & 1.29284  & 2.50093  & 3.34769  & 4.42284 \\ \hline
$\chi$=0.4   & 0  & $1.74\times10^{-5}$  & 0.001738  & 0.17561  & 0.72488  & 1.72374  & 3.33447  & 4.46343  & 5.89691 \\ \hline
$\chi$=0.5   & 0  & $2.17\times10^{-5}$   & 0.002173  & 0.21950  & 0.90607  & 2.15461  & 4.16795  & 5.57910  & 7.37087 \\ \hline
\end{tabular}
\end{table}
%%%%%%%%%%%%%%%%%%%%%%%%%%%%%%%%%%%%%%%%%%%%%%%%%%%%%%%%%%%%%%%%%%%%%%%%%%%%
\section{Parameter Estimation}\label{sec=params}

\subsection{Waveforms}
We can obtain the inspiral trajectory from adiabatic evolution in Eq.\eqref{eq:orbittime}, then it is convenient to compute the GW waveforms in the quadrupole approximation. The metric perturbation in the transverse-traceless(TT) gauge is
\begin{equation}
h_{i j}^{\mathrm{TT}}=\frac{2}{d_L}\left(P_{i l} P_{j m}-\frac{1}{2} P_{i j} P_{l m}\right) \ddot{I}_{l m}
\end{equation}
here $d_L$ is the source luminosity distance, $P_{ij}=\delta_{ij}-n_i n_j$ is the projection operator onto the wave unit direction $n_j$, where $\delta_{ij}$ is the Kronecker delta function. $\ddot{I}_{ij}$ is the second time derivative of the mass quadrupole moment which is given in terms of the source stress-energy tensor
\begin{equation}
I_{i j}=\int d^{3} x T_p^{t t}\left(t, x^{i}\right) x^{i} x^{j}=m_p y_p^{i}(t) y_p^{j}(t)
\end{equation}
where stress-energy tensor component $T_p^{tt}$ can be seen in Eq.\eqref{eq:sourceT}. So the strain produced by the GW and measured by the detector is then given by
\begin{equation}\label{eq:signal}
h(t)=h_{+}(t) F^{+}(t)+h_{\times}(t) F^{\times}(t)
\end{equation}
where $h_+(t)=\mathcal{A}\cos\left[2\varphi_{\rm orb}+2\varphi_0\right]\left(1+\cos^2\iota\right)$, $h_\times(t)=-2\mathcal{A}\sin\left[2\varphi_{\rm orb}+2\varphi_0\right]\cos\iota$,  $\iota$ is the inclination angle between the binary orbital angular momentum and the line of sight, and the GW amplitude  $\mathcal{A}=2m_{\rm p}\left[M\Omega(t)\right]^{2/3}/d_L$, $d_L$ is the luminosity distance.
The interferometer pattern functions $F^{+,\times}(t)$ and $\iota$ can be expressed in terms of four angles which specify the source orientation, $(\theta_s,\phi_s)$, and the orbital angular direction $(\theta_1,\phi_1)$.

%%%%%%%%%%%%%%%%%%%%%%%%%%%%%%%%%%%%%%%%%%%%%%%%%%%%%%%%%%%%%%%%%%%%%%%%%%%%
\subsection{Faithfulness}
In the time domain, we can use twelve parameters
\begin{equation}
\vec{\xi}=(ln~M, ln~m_p, a, \chi, d, r_0, \varphi_0, \theta_s, \phi_s, \theta_1, \phi_1, d_L)\nonumber
\end{equation}
to determine the GW waveform \eqref{eq:signal}. Here we fix the source angles $\theta_s=\pi/3,~\phi_s=\pi/2$ and $\theta_1=\phi_1=\pi/4$, the initial phase is set as $\varphi_0=0$ and the initial orbital separation is set to $r_{start}=11.53M$ and consider one-year adiabatic evolution before plunge $r_{ISCO}$. As mentioned above, we consider the model $m_p=10M_{\odot}$, $M=4\times10^5M_{\odot}$ with $a=0.9M$, here we choose $d=0.5$ to vary the secondary spin $\chi$. And the luminosity distance $d_L$ is a free scale factor for $h(t)$.

Introducing the noise-weighted inner product between two templates
\begin{equation}\label{eq:product}
\left\langle h_{1} \mid h_{2}\right\rangle=4 \Re \int_{f_{\min }}^{f_{\max }} \frac{\tilde{h}_{1}(f) \tilde{h}_{2}^{*}(f)}{S_{n}(f)} d f,
\end{equation}
$\tilde{h}_{1}(f)$ is the Fourier transform of the time domain signal, while its complex conjugate is $\tilde{h}_{2}^{*}(f)$. And $S_n(f)$ is the noise spectral density, which will be given for LISA, Taiji and TianQin respectively in the next subsection.

Noticed that the signal \eqref{eq:signal} is sampled in the time domain, which will be applied by a discrete Fourier transform evaluating the integral \eqref{eq:product} between $f_{min}=10^{-4}$Hz and $f_{max}=f_{N}$ here $f_N$ is the Nyquist frequency. The component related to the latter has been set to zero, and only $f\geq f_{min}$ Fourier components have been included. Before passing to the frequency space we taper $h(t)$ with a Tukey window with cosine fraction $\tau = 0.05$. The signal-to-noise ratio(SNR) can be obtained by $\rho=\left\langle h|h \right\rangle^{1/2}$. The Faithfulness between two GW signals is determined by
\begin{equation}
\mathcal{F}\left[h_1, h_2\right]=\max _{\left\{t_c, \phi_c\right\}} \frac{\left\langle h_1 \mid h_2\right\rangle}{\sqrt{\left\langle h_1 \mid h_1\right\rangle\left\langle h_2 \mid h_2\right\rangle}}
\end{equation}
where $(t_c, \varphi_c)$ are time and phase offsets.

%%%%%%%%%%%%%%%%%%%%%%%%%%%%%%%%%%%%%%%%%%%%%%%%%%%%%%%%%%%%%%%%%%%%%%%%%%%%
\subsection{Detector Configurations}
 As discussed above, the GW strain signal detected in a space-based GW detector is shown in Eq.\eqref{eq:signal}, where the antenna pattern functions $F^{+,\times}$ describing the detector response to sources with different locations and polarizations, are given by
\begin{equation}
\begin{aligned}
&F^{+}_1\left(\theta_{s}, \phi_{s}, \psi_{s}\right)=\frac{\sqrt{3}}{2}\left[\frac{1}{2}\left(1+\cos ^{2} \theta_{s}\right) \cos 2 \phi_{s} \cos 2 \psi_{s}-\cos \theta_{S} \sin 2 \phi_{s} \sin 2 \psi_{s}\right], \\
&F^{\times}_1\left(\theta_{s}, \phi_{s}, \psi_{s}\right)=\frac{\sqrt{3}}{2}\left[\frac{1}{2}\left(1+\cos ^{2} \theta_{s}\right) \cos 2 \phi_{s} \sin 2 \psi_{s}+\cos \theta_{s} \sin 2 \phi_{s} \cos 2 \psi_{s}\right],
\end{aligned}
\end{equation}
and the antenna pattern function of the second orthogonal Michelson interferometer can be written as
\begin{equation}
\begin{aligned}
&F_{\mathrm{2}}^{+}\left(\theta_{s}, \phi_{s}, \psi_{s}\right)=F_{\mathrm{1}}^{+}\left(\theta_{s}, \phi_{s}-\frac{\pi}{4}, \psi_{s}\right), \\
&F_{\mathrm{2}}^{\times}\left(\theta_{s}, \phi_{s}, \psi_{s}\right)=F_{\mathrm{1}}^{\times}\left(\theta_{s}, \phi_{s}-\frac{\pi}{4}, \psi_{s}\right).
\end{aligned}
\end{equation}
here $(\theta_s,\phi_s)$ describe the source location of EMRIs in the sky and the polarization angle function can be expressed as
\begin{equation}
\tan \psi_{S}=\frac{\hat{L} \cdot \hat{z}-(\hat{L} \cdot \hat{N})(\hat{z} \cdot \hat{N})}{\hat{N} \cdot(\hat{L} \times \hat{z})}
\end{equation}
where $\hat{L}$ and $-\hat{N}$ are the unit vector along the orbital angular momentum and the direction of GW propagation, respectively. Then, the Doppler phase due to the detector's orbital motion into the GW signal is
\begin{equation}
\varphi_{\text {orb}}(t) \rightarrow \varphi_{\text {orb }}(t)+\frac{d \varphi_{\text {orb }}(t)}{d t} R_{\mathrm{AU}} \sin \theta_{s} \cos \left(2 \pi t /(1 \text { year })-\phi_{s}\right).
\end{equation}

Now we introduce the LISA and Taiji power spectral density(PSD) which consists of the instrumental and confusion noise produced by unresolved galactic binaries \cite{Kang:2021bmp,Robson:2018ifk}
\begin{equation}
\begin{aligned}
S_{n}^{LISA,Taiji}(f)= \frac{10}{3 (L^{L,T})^{2}}\left[P_{\mathrm{OMS}}^{LISA,Taiji}+2\left(1+\cos ^{2}\left(f / f_{*}\right)\right) \frac{P_{\mathrm{acc}}}{(2 \pi f)^{4}}\right] \times\left[1+\frac{6}{10}\left(\frac{f}{f_{*}}\right)^{2}\right],
\end{aligned}
\end{equation}
where $f_*=c/(2\pi L^{LISA,Taiji} )$ is the transfer frequency and the arm length of space-borne GW detector is given by LISA $L^{L}=2.5\times10^6$km, Taiji $L^{T}=3\times10^6$km. And we have
\begin{eqnarray}
	P_{\mathrm{acc}}=(3 \times 10^{-15} \mathrm{m} / \mathrm{s}^{2})^2 \left[1+\left(\frac{0.4 \mathrm{mHz}}{f}\right)^{2}\right] \left[1+\left(\frac{f}{8 \mathrm{mHz}}\right)^{4}\right]\mathrm{Hz}^{-1}
\end{eqnarray}
\begin{equation}
\begin{aligned}
P_{\mathrm{OMS}}^{\mathrm{LISA}}&=&(1.5 \times 10^{-11} \mathrm{m})^2 \left[1+\left(\frac{2 \mathrm{mHz}}{f}\right)^{4}\right]\mathrm{Hz}^{-1}, \\
P_{\mathrm{op}}^{\mathrm{Taiji}}&=&(8 \times 10^{-12} \mathrm{m})^2 \left[1+\left(\frac{2 \mathrm{mHz}}{f}\right)^{4}\right] \mathrm{Hz}^{-1}.
\end{aligned}
\end{equation}

The sky averaged sensitivity of TianQin is given by \cite{Wang:2019ryf}
\begin{equation}
\begin{aligned}
&S_{n}^{TianQin}(f)=\frac{S_{N}^{TianQiin}(f)}{\bar{R}(2 \pi f)} \\
&S_{N}(f)=\frac{1}{(L^{TQ})^{2}}\left[\frac{4 S_{a}}{(2 \pi f)^{4}}\left(1+\frac{10^{-4} \mathrm{~Hz}}{f}\right)+S_{x}\right] \\
&\bar{R}(w)=\frac{3}{10} \times \frac{g(w \tau)}{1+0.6(w \tau)^{2}}
\end{aligned}
\end{equation}
where $L^{TQ}=\sqrt{3}\times10^5$km, $S_{a}^{1 / 2}=1 \times 10^{-15} \mathrm{~m} \mathrm{~s}^{-2} \mathrm{~Hz}^{-1 / 2}, S_{x}^{1 / 2}=1 \times 10^{-12} \mathrm{~m} \mathrm{~Hz}^{-1 / 2}$. $\tau=L^{TQ}/c$ is the travel time for TianQin arm length and
\begin{equation}
g(x)=\left\{\begin{array}{ccc}
\sum_{i=0}^{11} a_{i} x^{i} & : \quad x<4.1 \\
\exp [-0.322 \sin (2 x-4.712)+0.078] & : 4.1 \leq x<\frac{20 \pi}{\sqrt{3}}
\end{array}\right.
\end{equation}
with $(a_0,a_1,a_2,a_3,a_4,a_5,a_6,a_7,a_8,a_9,a_{10},a_{11})=(1,10^{-4},0.2639,4.62\times10^{-3},-0.16744,2.173\times10^{-2},2.101\times10^{-3},1.5135\times10^{-2},-8.4746\times10^{-3},1.76087\times10^{-3},-1.6046\times10^{-4},5.169\times10^{-6})$.

\clearpage

\bibliographystyle{jhep}
\bibliography{spinemri}
\end{document}